\documentclass[fleqn,usenatbib]{mnras}

\usepackage{newtxtext,newtxmath}
\usepackage[T1]{fontenc}

\DeclareRobustCommand{\VAN}[3]{#2}
\let\VANthebibliography\thebibliography
\def\thebibliography{\DeclareRobustCommand{\VAN}[3]{##3}\VANthebibliography}

\usepackage{graphicx}	% Including figure files
\usepackage{amsmath}	% Advanced maths commands
\usepackage{xcolor}

\title[Evolved massive stars in M31 and M33]{Near-infrared characterization of evolved massive stars in M31 and M33}

\author[M. Kraus et al.]{Michaela Kraus,$^{1}$\thanks{E-mail: michaela.kraus@asu.cas.cz (MK)}
Mar\'{i}a Laura Arias,$^{2,3}$
Michalis Kourniotis,$^{1}$
Andrea Torres,$^{2,3}$
Lydia S. Cidale,$^{2,3}$
\newauthor and Marcelo Borges Fernandes $^{4}$
\\
$^{1}$Astronomical Institute, Czech Academy of Sciences, Fri\v{c}ova 298, 251\,65 Ond\v{r}ejov, Czech Republic\\
$^{2}$Departamento de Espectroscop\'ia, Facultad de Ciencias Astron\'omicas y Geof\'isicas, Universidad Nacional de La Plata, Paseo del Bosque S/N, La Plata, \\ B1900FWA, Buenos Aires, Argentina\\
$^{3}$Instituto de Astrof\'isica de La Plata (CCT La Plata - CONICET, UNLP), Paseo del Bosque S/N, La Plata, B1900FWA, Buenos Aires, Argentina\\
$^{4}$Observat\'orio Nacional, Rua General Jos\'e Cristino 77, 20921-400 S\~ao Cristov\~ao, Rio 
de Janeiro, Brazil
}

\date{Accepted XXX. Received YYY; in original form ZZZ}

\pubyear{2025}

\begin{document}
\label{firstpage}
\pagerange{\pageref{firstpage}--\pageref{lastpage}}
\maketitle

% It should be a single paragraph not more than 250 words.
\begin{abstract}
The upper region of the Hertzsprung-Russell diagram is populated by massive stars in a diversity of evolutionary stages, and the classification of these stars is often based on observed characteristics exclusively in the optical spectral range. The near-infrared regime provides useful complementary information that can help resolving ambiguities in stellar classification and add valuable information about circumstellar envelopes or late-type companions. We present new, near-infrared medium-resolution $K$-band spectra for a sample of six evolved massive stars, four in M31 and two in M33. The spectra are obtained with the Gemini Near-Infrared Spectrograph (GNIRS) at the Gemini North telescope. We detect CO band emission from the environment of two M31 objects, J004320.97+414039.6 and J004621.08+421308.2, which we classify as B[e] supergiants, with J004320.97+414039.6 being most likely in a post-red supergiant stage. Two objects have pure emission from the hydrogen Pfund series. Of these, we propose that J004415.00+420156.2 in M31 could also be a B[e] supergiant while J013410.93+303437.6 (Var~83) is a well-known luminous blue variable (LBV) in M33. The M31 star J004229.87+410551.8 has a featureless spectrum and its evolutionary stage remains inconclusive; it could be an LBV undergoing an S~Dor cycle. The object J013242.26+302114.1 in M33 displays a pure absorption spectrum, including CO bands, consistent with its identification as a cool star. Radial velocity measurements of this red component, combined with modelling of the spectral energy distribution, suggest that J013242.26+302114.1 may  be a binary system consisting of an LBV or B[e] supergiant primary and a red supergiant secondary. If confirmed, it would represent the first of its kind.

\end{abstract}

\begin{keywords}
stars: emission-line, Be -- supergiants -- stars: winds, outflows -- circumstellar matter -- methods: observational -- techniques: spectroscopic
\end{keywords}

%%%%%%%%%%%%%%%%%%%%%%%%%%%%%%%%%%%%%%%%%%%%%%%%%%

%%%%%%%%%%%%%%%%% BODY OF PAPER %%%%%%%%%%%%%%%%%%
\section{Introduction}

Massive stars ($> 8$\,M$_{\odot}$) are the cornerstone to the dynamical and chemical 
evolution of their host galaxies. With their intense winds throughout the entire evolution 
and outburst activities in advanced evolutionary stages, massive stars enrich their 
environments with large quantities of chemically processed material and inject vast amounts 
of energy into them \citep[e.g.,][]{2022ARA&A..60..455E}. 

The upper part of the Hertzsprung-Russell (HR) diagram is populated with a plethora of objects 
displaying a variety of observational characteristics. In the blue region, we 
find classical blue supergiants (BSGs) along with B[e] supergiants (B[e]SGs) and Luminous 
Blue Variables (LBVs) in their quiescent state. Towards lower temperatures reside A- to 
G-type supergiants (YSGs) together with warm and yellow hypergiants (YHGs) as well as
LBVs undergoing an S~Dor cycle (also often referred to as "LBV outburst"). The 
cool end of the HR diagram of massive stars is the home of the red supergiants (RSGs), 
but also YHGs that are  
in outburst can dwell in this temperature domain. To add to the zoo of evolved massive stars, the terminology of `iron stars' or Fe\,{\sc ii} stars has been used in the literature to describe luminous hot stars that display a large amount of emission lines, predominantly Fe\,{\sc ii} and [Fe\,{\sc ii}], but often lack atmospheric absorption lines that would allow to classify the nature of the underlying star \citep{2000PASP..112...50W, 2012A&A...541A.146C, 2014ApJ...790...48H}. Such iron emission dominated spectra might represent LBVs in outburst but are also common for B[e]SGs. 

To tell all these objects apart is not 
always an easy task, as their characteristics can change considerably during outburst 
phases, mimicking (some of) the properties of other object classes, and a lot of effort 
has been undertaken to define robust classification criteria for each object class 
(see, e.g., \citeauthor{1994PASP..106.1025H} \citeyear{1994PASP..106.1025H} and 
\citeauthor{2020Galax...8...20W} \citeyear{2020Galax...8...20W} for LBVs, 
\citeauthor{1998A&A...340..117L} \citeyear{1998A&A...340..117L} and 
\citeauthor{2019Galax...7...83K} \citeyear{2019Galax...7...83K} for B[e]SGs, 
\citeauthor{1998A&ARv...8..145D} \citeyear{1998A&ARv...8..145D} and
\citeauthor{2009ASPC..412...17O} \citeyear{2009ASPC..412...17O} for YHGs, and 
\citeauthor{2013ApJ...773...46H} \citeyear{2013ApJ...773...46H} for warm 
hypergiants).

\begin{table*}
	\centering
	\caption{Objects and proposed classifications from literature and ours.}
	\label{tab:objects}
\begin{tabular}{llcccccc}
\hline
Galaxy & Object & Alternative       & $V$   & $K$   & Literature     & Reference & New \\
       & LGGS   & SIMBAD identifier & [mag] & [mag] & Classification &           & Classification \\
\hline
M31 & J004229.87+410551.8 & -- &  18.785 & 13.90 & cLBV\,/\,FeII\,/\,B[e]SG & 1\,/\,2\,/\,3 & \bf LBV in S~Dor cycle?  \\ 
M31 & J004415.00+420156.2 & BA 1-621 & 18.291 &  14.54 & cLBV\,/\,FeII\,/\,B[e]SG & 1\,/\,2\,/\,3 & {\bf B[e]SG}  \\  
M31 & J004320.97+414039.6 & BA 1-558 & 19.215 &  15.20 & cLBV\,/\,FeII\,/\,B[e]SG & 4\,/\,2\,/\,3 &  {\bf post-RSG B[e]SG}  \\  
M31 & J004621.08+421308.2 & [WB92a] 895 & 18.155 &  15.15 & cLBV\,/\,warm hypergiant & 4\,/\,5 & {\bf B[e]SG}  \\   
\hline 
M33 & J013242.26+302114.1 & IFM-B 57  & 17.440 &  13.88 & cLBV\,/\,FeII\,/\,B[e]SG & 1\,/\,6\,/\,3 &  \bf cLBV/B[e]SG + RSG?  \\  
M33 & J013410.93+303437.6 & IFM-B 1588, 	VHK 83$^{a}$ & 16.027 & 15.26 & LBV & 7 & {\bf LBV}  \\ 
\hline
\multicolumn{8}{l}{$^{a}$ formerly Var 83}\\
\multicolumn{8}{l}{$V$-band and $K$-band magnitudes are from \protect{\citet{2006AJ....131.2478M}} and the 2MASS survey and catalog \protect{\citep{2003yCat.2246....0C, 2006AJ....131.1163S}}, respectively,} \\ 
\multicolumn{8}{l}{except for the $K$-band magnitude for J004229.87+410551.8, which is from \protect{\citet{2010MNRAS.402..803P}}.}\\
\multicolumn{8}{l}{References: 1 - \protect{\citet{2007AJ....134.2474M}}, 2 - 
\protect{\citet{2014ApJ...790...48H}}, 3 - \protect{\citet{2017ApJ...836...64H}}, 4 - 
\protect{\citet{1998ApJ...507..210K}}, 5 - \protect{\citet{2016ApJ...825...50G}}, }\\
\multicolumn{8}{l}{6 - \protect{\citet{2012A&A...541A.146C}}, 7 - \protect{\citet{1975ApJS...29..303V}}.}
\end{tabular}

\end{table*}

However, the evolutionary paths of massive stars from the main-sequence to their deaths are yet most uncertain. Stellar evolution
calculations critically depend on a variety of input parameters concerning the internal structure and physical processes, such as mixing and core-overshooting, metallicity, initial rotation, and the star's mass-loss behaviour along the entire evolution \citep{2000ApJ...528..368H, 2000ApJ...544.1016H, 2012A&A...537A.146E, 2014MNRAS.439L...6G, 2014A&A...564A..30G, 2022A&A...658A.125S}. Many of these parameters cannot be (easily) accessed from observations, and inaccurate 
determinations of input parameter values can lead to huge uncertainties in the predictions from
evolutionary models \citep{2013A&A...560A..16M}.
Furthermore, possible or likely evolutionary connections between, for example, YHGs, warm hypergiants, B[e]SGs and LBVs are not yet fully established. It is also still unclear to what extent binary interaction affects the observed phenomena and, despite some progress made
\citep{2019Galax...7...83K, 2022A&A...657A...4M, 2024ARA&A..62...21M, 2025MNRAS.540L..28K}, there is still insufficient information on the multiplicity within the individual classes. Nevertheless, observations of the diverse objects help characterizing their class and evolutionary state. Discoveries of new members lead to more profound insight on massive star populations in these extreme evolutionary transition phases and on the evolutionary and dynamically history of massive stars in different host galaxies.   

The luminous stellar population of a galaxy sticks out due to the high optical brightness 
of its members, which makes it easier to identify them, and which makes studies of 
extra-galactic objects highly attractive. In the past decades, many surveys were dedicated to resolve the diverse massive star populations in galaxies within and beyond the Local Group by means of optical spectroscopy  \citep{1990AJ.....99...84H, 1996AJ....112.1450C, 1999A&AS..140..309F, 2005A&A...437..217F, 2009MNRAS.396L..21V, 2012A&A...541A.146C, 2013ApJ...773...46H, 2014ApJ...790...48H, 2016ApJ...825...50G, 2017ApJ...836...64H, 2017ApJ...844...40H, 2017A&A...601A..76K, 2019MNRAS.484L..24S, 2023ApJ...959..102D, 2024arXiv240615270S}, including tools based on machine learning \citep{2022A&A...666A.122M}, or by optical spectroscopy combined with multi-band photometry \citep{1996ApJ...469..629M, 2007AJ....134.2474M, 2014ApJ...788...83M, 2016AJ....152...62M, 2019AJ....157...22H, 2023Galax..11...79M, 2024A&A...686A..77B}. 
The identified objects were classified based on common 
characteristics and similarities with representatives of their classes in the Milky Way.

However, the classification of massive stars based solely on optical spectroscopy and 
multi-band photometry bears ambiguities, because stars in different evolutionary states 
can display very similar spectra with only subtle differences. In addition, an infrared excess 
detected in the spectral energy distribution can be caused by either circumstellar dust
or a late-type companion (or a combination of both). Therefore, complementary information 
is needed to unambiguously classify a star and its environment. 

A number of surveys of evolved massive stars have been carried out in the near-infrared spectral 
range revealing a great diversity of spectroscopic features, including emission from warm 
molecular gas from circumstellar environments \citep{1988ApJ...324.1071M, 1989A&A...223..237M, 
1988ApJ...334..639M, 1996ApJS..107..281H, 1996ApJ...470..597M}. Follow-up investigations by 
\citet{2013A&A...558A..17O} were dedicated to specifically unveil the near-infrared properties of 
YHGs, LBVs, and B[e]SGs, and to characterize their envelopes and (where 
appropriate) circumstellar disks. This study provided complementary information to discriminate 
between LBV candidates (cLBV) and B[e]SGs, as these share many characteristics at optical 
wavelengths. Combining the characteristics of these two classes of evolved massive stars from 
the optical and infrared spectral regions with information about their infrared colours 
resulted in a more reliable classification of objects \citep{2014ApJ...780L..10K, 
2019Galax...7...83K} and the identification of new candidates to both classes of objects 
\citep[e.g.,][]{2015MNRAS.447.2459S, 2018MNRAS.480.3706K, 2020AJ....160..166C}. 

Motivated by this, we started to systematically observe extra-galactic luminous stars with 
ambiguous classification in the near-infrared spectral domain with the aim to improve the 
classification of the objects and to study their environments. To achieve this, we mainly focus 
on detecting emission from the CO first overtone bands. These molecular bands are an immediate 
indication for a dense and warm ($< 5000$\,K) circumstellar environment 
\citep{1988ApJ...324.1071M, 2009A&A...494..253K}. Detection of emission from the CO bands serves 
as a complementary criterion to classify a star as B[e]SG \citep{2014ApJ...780L..10K} rather 
than as cLBV, although we note that only about 50\% of the B[e]SGs show these bands 
\citep{2012MNRAS.426L..56O, 2013A&A...558A..17O, 2019Galax...7...83K}. Conversely, the 
absence of CO band emission does not automatically exclude the B[e]SG nature of the 
object. If detected, the CO bands have a further advantage, because they consist of 
emission from the isotopes $^{12}$CO and $^{13}$CO, and their ratio indicates whether the 
star was already evolved or yet unevolved at the time of ejection of the material  
\citep{2009A&A...494..253K, 2010MNRAS.408L...6L}.

In this work, we present medium-resolution K-band spectra of six objects listed in
Table~\ref{tab:objects}, one being a proposed LBV star in M33. Five of the objects have literature 
classification as either cLBV, iron star (FeII), B[e]SG, or warm hypergiant, based on studies of 
their optical spectral characteristics, photometric variability, and spectral energy distribution. 
Four of these objects reside in M31 and one in M33. For four objects, these are the first 
near-infrared spectra.

\begin{figure*}
\includegraphics[width=\textwidth]{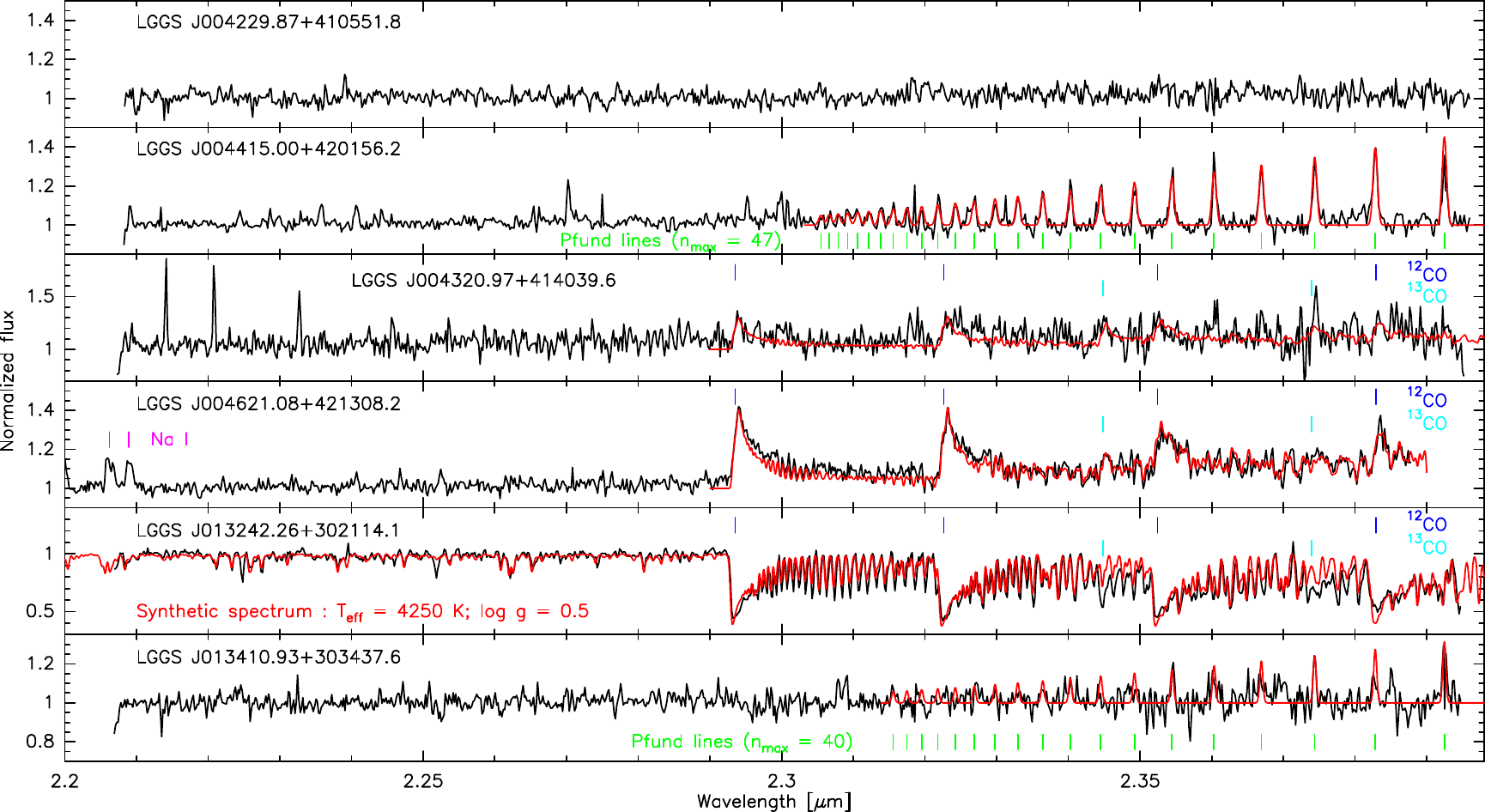}
\caption{Best-fitting models (red) to the circumstellar emission and absorption seen in the normalized GNIRS spectra (black) of our objects. The parameters for the CO band and Pfund line emission are listed in Tables~\ref{tab:CO} and \ref{tab:Pf}, respectively. The parameters for the atmospheric model are listed in the corresponding panel. The order of the objects follows the one in Table~\ref{tab:objects}.
\label{fig:Kband}}
\end{figure*}

\section{Observations and data reduction}
\label{sec:obs} % used for referring to this section from elsewhere

Medium-resolution ($R \sim 5900$) $K$-band spectra of our objects were obtained with the 
Gemini Near-InfraRed Spectrograph \citep[GNIRS,][]{2006SPIE.6269E..14E, 
2006SPIE.6269E..4CE} attached to the 8.1-m telescope at Gemini North Observatory. 
Data acquisition was 
between January 2019 and September 2020 with the short camera ($0\farcs15$\,pixel$^{-1}$), 
a slit width of $0\farcs3$, and the $111$~l/mm grating 
centred at $2.3\,\mu$m for a wavelength coverage of $2.2\,\mu$m $-\,2.4\,\mu$m. The 
observations were carried out in AB pairs to facilitate sky subtraction. Data reduction 
was performed using standard {\sc IRAF}\footnote{IRAF is distributed by the National 
Optical Astronomy Observatory, which is operated by the Association of Universities for 
Research in Astronomy (AURA) under cooperative agreement with the National Science 
Foundation.} tasks. The individual steps include flat-fielding, wavelength calibration, 
and removal of telluric features. 

To perform the telluric correction, spatially close-by late B- or early A-type 
main-sequence stars were observed 
immediately before or after the target exposure at similar airmass. Choosing telluric 
standard stars in this narrow spectral-type range has the advantage that the telluric star itself 
has no intrinsic spectral features in the observed spectral region ($2.2-2.4\,\mu$m). Their 
spectra hence serve as perfect template for correcting the science spectra from telluric 
absorption features. The selected telluric stars are also bright ($K = 6-7$\,mag), allowing 
to obtain a spectrum with significant signal-to-noise (S/N) ratios (typically $100-120$) with 
short exposures. The telluric correction was performed with the {\sc IRAF} task {\sl telluric}, 
which enables interactive scaling and wavelength shifting of the telluric standard spectrum to 
achieve optimal cancellation of the atmospheric (telluric) features in the target spectrum.

Finally, the spectra were normalized by fitting the continuum with low-order polynomial.
S/N ratios of the spectra are 33 (J004229.87+410551.8), 90 (J004415.00+420156.2), 
16 (J004320.97+414039.6), 52 (J004621.08+421308.2), 82 (J013242.26+302114.1), 
%52 (J013350.12+304126.6), 
and 28 (J013410.93+303437.6). 

Details on our targets are listed in Table~\ref{tab:objects}, where we provide the name 
of the host galaxy, the identifier of the object from the Local Group Galaxies Survey 
\citep[LGGS,][]{2006AJ....131.2478M} along with their archival $V$- and $K$-band 
photometry and previous classifications found in the literature.

\section{Results} 
\label{sec:res}

The normalized spectra, shifted to the rest wavelength\footnote{except for J004229.87+410551.8 because the spectrum contains no reference line to determine the shift.}, are shown in black in Fig.~\ref{fig:Kband}. We detect CO band 
emission from J004320.97+414039.6 and J004621.08+421308.2, CO band absorption 
in J013242.26+302114.1, emission from the hydrogen Pfund series 
from J004415.00+420156.2 and J013410.93+303437.6, while the spectrum of 
J004229.87+410551.8 appears to be featureless\footnote{The same telluric standard star spectrum has been used to clean the spectra of J004229.87+410551.8 and J004415.00+420156.2 (as it was observed in between the two target exposures) and in both science spectra the telluric features were equally good and completely removed. We also checked that the K-band spectrum is not contaminated by the near-by cluster \citep{2012ApJS..199...37K} as the cluster is well separated from the object on the acquisition image and does not fall into the slit. Hence, the featureless spectrum of J004229.87+410551.8 is real.}. In the following, we model the emission 
from the CO bands and from the Pfund series to derive the physical parameters 
(temperature, density, dynamics) of the circumstellar environment of these objects.

\begin{table*}
	\centering
	\caption{Best-fitting parameters of the CO band emission.}
	\label{tab:CO}
\begin{tabular}{|lccccc|}
\hline
Object	 & $T_{\rm CO}$	& $N_{\rm CO}$ & $^{12}{\rm CO}/^{13}{\rm CO}$ & $v_{\rm rot, los}$ & $v_{\rm Gauss}$ \\
LGGS  & [K] & [cm$^{-2}$] & & [km\,s$^{-1}$]  & [km\,s$^{-1}$]  \\
\hline
J004320.97+414039.6  & 2200$\pm$100 &  (3$\pm$1)  $\times$ 10$^{21}$   &  3$\pm$1  & 60$\pm$10  & 2$\pm$0.5 \\
J004621.08+421308.2  & 1500$\pm$100 &  (5$\pm$1)  $\times$ 10$^{22}$  & 50$\pm$10 & 60$\pm$10  & 2$\pm$0.5 \\
\hline
\end{tabular}
\end{table*}

\subsection{CO band emission}
\label{sec:CO}

The positions of the individual band heads of both $^{12}$CO and $^{13}$CO are indicated 
with vertical bars in Fig.~\ref{fig:Kband}. The observed shape of the CO band heads with a 
blue-shifted shoulder and a red-shifted maximum is characteristic of (Keplerian) rotation 
\citep{1995Ap&SS.224...25C, 2019Galax...7...83K} in a disk (or ring) revolving around the 
star. But it might also indicate equatorial outflow at constant velocity, which generates 
the same double-peaked profiles for the individual ro-vibrational lines as rotation 
\citep[e.g.,][]{2010A&A...517A..30K}. In 
either case, the shape of the band heads suggests that the CO gas around 
J004320.97+414039.6 and J004621.08+421308.2 is most likely concentrated in a ring around 
the object, similar to what is typically found for B[e]SGs \citep{2012A&A...548A..72C, 
2012ASPC..464...67M, 2015AJ....149...13M, 2013A&A...549A..28K,  
2023Galax..11...76K}. Furthermore, the analysis of multiple disk tracers demonstrated that 
B[e]SGs are surrounded by multiple rings of gas and dust in (quasi-) Keplerian rotation
\citep{2012MNRAS.423..284A, 2016A&A...593A.112K, 2018MNRAS.480..320M, 2018A&A...612A.113T}. 
Therefore, we assume that the CO gas around our targets is also rotating. 

\begin{figure}
\includegraphics[width=0.47\textwidth]{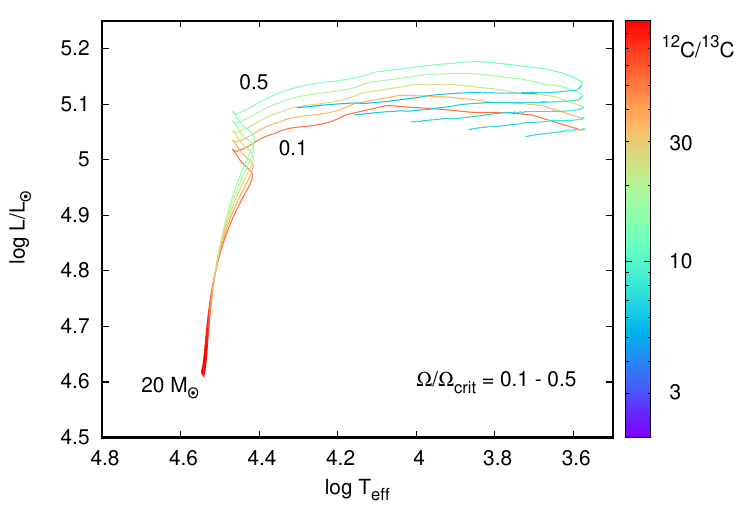}
\caption{Stellar evolution tracks for rotating stars with solar metallicity and initial mass of 20\,M$_{\odot}$. Tracks are interpolations of those computed by \citet{2012A&A...537A.146E}. The colour code represents the abundance ratio of the carbon isotopes during the evolution. For all rotation rates, values below 5 are clearly reached only after the stars have passed through the RSG evolution. 
\label{fig:C-ratio}}
\end{figure}

To generate 
synthetic CO band spectra, we utilize our code developed by \citet{2000A&A...362..158K} 
to compute the emission of $^{12}$CO from a Keplerian rotating disk, and advanced by 
\citet{2009A&A...494..253K} and \citet{2013A&A...558A..17O} by adding $^{13}$CO. In 
addition, we have implemented updated values for the energy levels and Einstein 
transition coefficients for both $^{12}$CO and $^{13}$CO from \citet{2015ApJS..216...15L}. 
The code computes the CO band emission under the assumption of LTE conditions. This is a 
reasonable assumption given the high column densities usually found in these molecular 
rings. The best-fitting parameters are listed in Table~\ref{tab:CO}, and the fits are 
included in Fig.~\ref{fig:Kband}. The parameters $T_{\rm CO}$ and $N_{\rm CO}$ denote the 
temperature and column density of the CO gas, and the gas dynamics is represented by the
rotation velocity projected to the line-of-sight, $v_{\rm rot, los}$, and a Gaussian 
component, $v_{\rm Gauss}$, combining the contributions from the thermal and turbulent 
motions. The value for the abundance ratio of the molecular isotopes $^{12}$CO/$^{13}$CO 
marks the amount of enrichment of the circumstellar environment in $^{13}$C. This value mirrors 
the stellar surface isotope abundance ratio $^{12}$C/$^{13}$C at the time of material ejection
from the star. The atomic isotopic abundance ratio has an initial (interstellar) value of about 90 
and gradually decreases during stellar evolution. This is shown in Fig.~\ref{fig:C-ratio} for 
evolutionary tracks of a star at solar metallicity (which is similar to the metallicity of M31) 
with an initial mass of 20\,M$_{\odot}$ and for rotation rates $\Omega/\Omega_{\rm crit} = 0.1 - 
0.5$ in steps of $0.1$ \citep{2012A&A...537A.146E}. The tracks are interpolations obtained by the 
tool SYCLIST\footnote{https://www.unige.ch/sciences/astro/evolution/en/database/syclist/} \citep{2022MNRAS.511.2814Y}. With 
increasing rotation velocity of the star, the enrichment in $^{13}$C on the stellar surface starts 
earlier in the evolution, but values of the abundance ratio below 5 occur only after the star 
has passed through the RSG phase. The shown tracks approximately represent the evolution of our 
two objects with CO band emission, for which luminosities of $\log L/L_{\odot} = 5.0-5.1$ have 
been determined \citep{2014ApJ...790...48H, 2016ApJ...825...50G}. The derived CO abundance ratios 
indicate that the environments of both stars are enriched in $^{13}$CO.

We emphasize that the parameters for J004320.97+414039.6 bear some uncertainty 
due to the very weak CO signal compared to the noise and significant remnants of telluric 
lines polluting the red portion of the spectrum. Nevertheless, the high enrichment in 
$^{13}$CO in the spectrum of this object appears to be real, and the low errorbar in the 
abundance ratio of the CO isotopes (despite the lower quality of the spectrum) is due to the 
significantly increased sensitivity of the emission spectrum with increasing $^{13}$CO abundance. 

In contrast, J004621.08+421308.2 is found to have a low $^{13}$CO abundance and the value of 
$^{12}$CO/$^{13}$CO is less well constrained, resulting in a considerably larger errorbar. On the 
other hand, the star's environment has a significantly higher CO column density than 
J004320.97+414039.6. This higher column density results in an increase in optical depth across the 
$^{12}$CO band heads and, consequently, in a reduced relative flux ratio between the $^{12}$CO and 
$^{13}$CO band heads. Therefore, the $^{13}$CO band heads become prominent in the observed 
spectrum despite the 50 times lower column density of the $^{13}$CO gas.

\subsection{Pfund line emission}
\label{sec:Pf}

In contrast to the CO bands, the Pfund lines detected in J004415.00+420156.2 and 
J013410.93+303437.6 display single-peaked profiles suggesting that the lines form in the 
ionized wind rather than in a rotating or outflowing disk. The maximum detectable Pfund 
line, $n_{\rm max}$, is an indicator for the hydrogen density, $N_{\rm H}$, in the 
line-forming region. We use the code developed by \cite{2000A&A...362..158K} which adopts 
Menzel's case B recombination, fix the electron temperature at $10\,000$\,K, 
which is a typical value in the winds of B-type supergiants\footnote{A slightly 
higher or lower electron temperature does not affect the resulting shape of the emission spectrum, 
just the total line intensities. But these differences disappear when normalizing the synthetic 
spectra.}, and assume that the lines are optically thin and their profile shapes can be 
represented with a Gaussian profile. The best-fitting parameters for the hydrogen density, 
maximum number of detected Pfund line, and kinematics are listed in Table~\ref{tab:Pf}, and the 
fits are included in Fig.~\ref{fig:Kband}. The parameters derived for J013410.93+303437.6 should 
be treated with caution because the spectrum exhibits a high noise level compared to the intensity 
of the emission lines and some remnants of telluric lines are present in the red part of the 
spectrum. Due to the high noise level, the maximum detectable Pfund line might be higher 
(lower). Consequently, the hydrogen density within the line forming region might be a lower 
(upper) limit.

\begin{table}
	\centering
	\caption{Best-fitting parameters for the emission of the Pfund series.}
	\label{tab:Pf}
\begin{tabular}{lccc}
\hline
Object	 &  $N_{\rm H}$ & $n_{\rm max}$ & $v_{\rm Gauss}$  \\
LGGS  &  [cm$^{-3}$] & & [km\,s$^{-1}$]  \\
\hline
J004415.00+420156.2  &  (1.9$\pm$0.3) $\times$ 10$^{13}$ &  47 & 60$\pm$10  \\
J013410.93+303437.6  &  (4.9$\pm$0.7) $\times$ 10$^{13}$ &  40 & 60$\pm$10  \\
\hline
\end{tabular}
\end{table}

\subsection{Absorption spectra}
\label{sec:companion}

The object J013242.26+302114.1 presents a pure absorption spectrum with very intense CO bands in 
the portion of the K-band region covered by our observations. CO bands in absorption are 
typically seen from the photospheres of late-type stars, which could mean that a cool object 
appears on the same sky position as the hot supergiant star 
J013242.26+302114.1\footnote{Our acquisition image 
showed only a single point-source object at the position of J013242.26+302114.1.}. To verify
this, we computed models with the spectrum synthesis code Turbospectrum using MARCS 
atmospheric models \citep{2012ascl.soft05004P}. Unfortunately, these atmosphere models 
contain only $^{12}$CO whereas the observed spectrum shows clear indication for intense 
$^{13}$CO as well. Therefore, we include in Fig.~\ref{fig:Kband} a synthetic spectrum for 
pure demonstration purpose, not with the aim to have the best-fitting model. We find that the 
general spectral features are decently well represented with an effective temperature between 
4000 and 4250\,K and a $\log g$ in the range $0.5-1.0$. However, these temperature values should 
be regarded as upper limit, because any contribution to the near-infrared continuum emission from 
either a hot component or circumstellar dust reduces the observable line intensities, meaning that 
the real temperature of the cool component will be lower.

\section{Discussion}

The $K$-band spectra of the six studied objects display a large diversity. Nevertheless,
when combining the new information with what is known about the objects from the 
literature, they help address the question of classification, at least for those stars
with distinct emission features. 

\subsection{Objects in M31}

\subsubsection{An LBV in an S~Dor cycle?}
The object J004229.87+410551.8 has been classified by \citet{2007AJ....134.2474M} as a 
hot cLBV. According to their definition, a hot cLBV is a star with an optical spectrum 
resembling the one of an LBV in its visual minimum and thus its quiescent S Dor phase. This 
means that the optical spectrum displays intense emission of the Balmer lines and of 
singly ionized metals, primarily [Fe\,{\sc ii}]. In contrast, \citet{2014ApJ...790...48H}
used the numerous iron emission lines as a criterion to classify the object as a Fe\,{\sc ii} 
star, whereas later-on \citet{2017ApJ...844...40H} proposed a B[e]SG status due to the infrared 
excess seen with data from \textit{Spitzer} and WISE. However, the star is in a crowded region 
\citep{2016AJ....152...62M}, and as mentioned by \citet{2014ApJ...790...48H}, a cluster is 
at only $1\arcsec$ separation to J004229.87+410551.8. Because both Spitzer IRAC and WISE 
have low angular resolution ($2\arcsec$ and more than $6\arcsec$, respectively), a 
contamination of the stellar infrared fluxes cannot be excluded. Based on this, a B[e]SG 
status of J004229.87+410551.8 has already been questioned by \citet{2019Galax...7...83K}. 
Our featureless $K$-band spectrum adds further evidence that the star is not a classical 
B[e]SG, because B[e]SGs typically display either intense Pfund line emission, or CO band 
emission, or both. Galactic as well as Magellanic Cloud B[e]SGs with CO band emission  
populate the luminosity range $\log L/L_{\odot} \sim 5.0 - 5.9$ \citep{2013A&A...558A..17O,
2019Galax...7...83K, 2023Galax..11...76K}. With a luminosity of J004229.87+410551.8 of 
$\log L/L_{\odot} \simeq 5.2$ \citep{2014ApJ...790...48H}, the star falls well into this 
range. The absence of both Pfund line and CO band emission suggests that neither the 
wind nor the equatorial environment of the star are dense enough to create detectable emission.
Therefore, we tentatively exclude a B[e]SG classification of the object.

\begin{figure}[ht]
\includegraphics[width=0.47\textwidth]{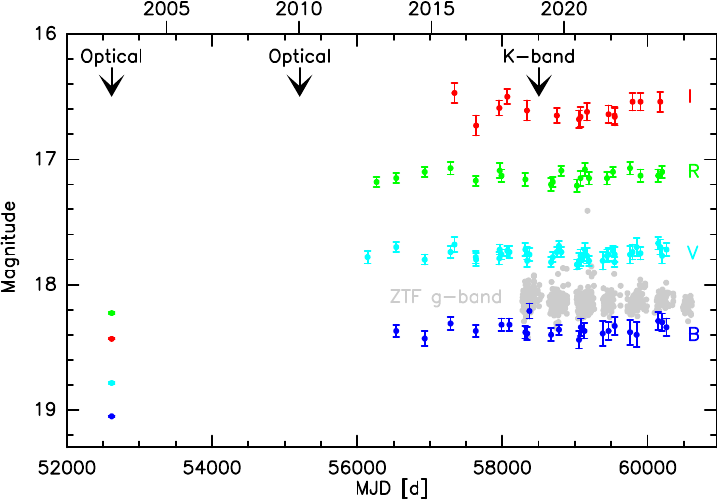}
\caption{Time series observations of J004229.87+410551.8 collected in $g$ band with ZTF (gray points) and $BVRI$ filters (coloured points). For clarity of the plot, the error bars of the ZTF measurements (which are of similar size as for the $BVRI$ photometry) are omitted. The errorbars of the data from 2002 are smaller than the symbol size. The black arrows mark the dates of spectroscopic observations in the optical \citep{2007AJ....134.2474M, 2014ApJ...790...48H} and in the K-band (ours).
\label{fig:ZTF}}
\end{figure}

We also consider the hypothesis that J004229.87+410551.8 might be an LBV candidate, as 
proposed by \citet{2007AJ....134.2474M}. The $K$-band survey of 
\citet{2013A&A...558A..17O} revealed that LBVs display intense Pfund line emission when they are 
in their hot, quiescent state, while they show a featureless (e.g. LHA 120-S 155) or an absorption 
spectrum mimicking a cool (A- to F-type) supergiant star (e.g. WRAY 15-751) when they are in outburst. 
The absence of Pfund line emission seems to speak against an LBV in quiescence. To check whether 
J004229.87+410551.8 could be in outburst, the star should show a brightening. The database of the 
Zwicky Transient Facility (ZTF) photometric survey \citep{2019PASP..131f8003B, 
2019PASP..131a8003M} provides observations of J004229.87+410551.8 through data release 
DR23\footnote{https://irsa.ipac.caltech.edu/Missions/ztf.html}. These observations were conducted 
in the optical $g$ band between June 2018 and October 2024. Furthermore, we inspected the 
catalogue of $BVRI$ photometric observations of luminous stars in M31 collected between August 
2012 and November 2023 at the UIS Henry Barber Research 
Observatory\footnote{https://uisacad5.uis.edu/~jmart5/M31M33photcat/}, of which the results from 
the first four years of monitoring have been published by \citet{2017AJ....154...81M}. 
Figure~\ref{fig:ZTF} shows the time series of the observations with the date of our observations 
marked. The photometric measurements in all bands display a scatter of about $0.2-0.3$\,mag around 
the average magnitude, which might indicate some low-amplitude quasi-periodic variability. 
We also added the photometric measurements of \cite{2006AJ....131.2478M} from September 
2002 and marked the dates of spectroscopic observations taken in September 2002 
\citep{2007AJ....134.2474M} and in October 2010 \citep{2014ApJ...790...48H}. These data clearly 
demonstrate that in 2002 the star was dimmer ($\sim 0.8$\,mag in $B$, $\sim 1$\,mag in $V$, 
$\sim 2$\,mag in $I$) and bluer ($B-V= 0.27$ in 2002 versus $0.62$ in 2023; $V-I = 0.35$ in 2002 
versus $1.18$ in 2023). Brightening and simultaneous reddening are typical characteristics of an 
S~Dor variability \citep{2020Galax...8...20W}. This suggests that J004229.87+410551.8 might have 
entered an S~Dor cycle somewhen between the last optical spectroscopic observation in 2010 and the 
beginning of the photometric monitoring in 2012 and that it is stable since then. The duration of 
S~Dor cycles can spread from years to decades \citep{2020Galax...8...20W}.  
To confirm or refute this and to further follow the evolution of this object, new optical spectroscopic observations combined with new multiband photometry are urgently needed.
 
\subsubsection{B[e] supergiants}

The objects J004415.00+420156.2 and J004320.97+414039.6 have previously also
been classified as (hot) cLBVs by \citet{2007AJ....134.2474M} and  
\citet{1998ApJ...507..210K}, as Fe\,{\sc ii} stars by \citet{2014ApJ...790...48H} and as 
B[e]SGs by \citet{2017ApJ...844...40H}. Likewise, a B[e]SG status was proposed for both 
objects by \citet{2019Galax...7...83K} due to their infrared colours, which clearly 
separate them from the region populated by LBVs in the colour-colour diagrams. Optical and 
near-infrared spectra of J004415.00+420156.2 have also been presented by 
\citet{2020MNRAS.497..687S} who found clear indication of free-free and free-bound 
emission from an ionized wind. From the presented data it is inconclusive whether emission 
from the Pfund series is seen in their spectrum, and the authors do not mention it. 
Our $K$-band spectrum of J004415.00+420156.2 displays only emission from the Pfund series. 
However, its high luminosity of $\log (L/L_{\odot}) = 5.73$ \citep{2014ApJ...790...48H} aligns 
the star with high-luminosity B[e]SGs (LHA 120-S 22, LHA 120-S 127) in the Large 
Magellanic Cloud (LMC) lacking CO band emission \citep{2019Galax...7...83K}. In contrast, 
J004320.97+414039.6 displays weak CO band emission. With its luminosity of $\log 
(L/L_{\odot}) = 5.01$ \citep{2014ApJ...790...48H}, it resides at the lower luminosity 
border of B[e]SGs showing CO band emission \citep{2019Galax...7...83K}. The strong 
enrichment in $^{13}$CO suggests an evolved, presumably post-red supergiant evolutionary 
state, but a better quality spectrum would be required to confirm this.

The last object in our M31 sample is J004621.08+421308.2. It was suggested to be 
a cLBV by \citet{1998ApJ...507..210K}, whereas \citet{2016ApJ...825...50G} proposed it 
could belong to the group of warm hypergiants with an infrared excess. While 
\citet{2016ApJ...825...50G} state that the stellar spectrum would indicate a late A 
spectral type due to the absence of He\,{\sc i} emission, they assign the star an 
effective temperature of $\sim 9000$\,K and a spectral type A0 in their table 4. 
Follow-up optical and near-infrared spectroscopic observations were carried out by 
\citet{2020MNRAS.497..687S}, who reported the detection of CO band emission from that 
star. Their spectrum has somewhat lower resolution than our GNIRS spectrum and covers only 
the first two band heads of $^{12}$CO, which makes it difficult to estimate the physical 
parameters of the CO emitting region. Nevertheless, the molecular emission seems to be 
persistent, as we also detected it in our spectrum taken about eight years later. While 
the CO temperature obtained from our modeling is lower than what is found in most disks around B[e]SGs, it is comparable to the Galactic B[e]SG star MWC\,349 \citep{2020MNRAS.493.4308K}. 
Intense and persistent CO band emission is characteristic for many B[e]SGs, while it is
only occasionally seen in YHGs \citep[see][for an 
overview]{2023Galax..11...76K}. The revised stellar temperature $10\,000-15\,000$\,K 
by \citet{2020MNRAS.497..687S} from both their spectrum (that shows He\,{\sc i} emission) and 
spectral energy distribution (SED) modelling, agrees with a spectral-type B so that we tentatively classify the star as a 
B[e]SG. The rather low enrichment in $^{13}$CO seen in our spectrum speaks in favour of a 
star that is only slightly evolved (pre-RSG) at the time of material ejection. 
The small wavelength shift of our spectrum of J004621.08+421308.2 allows us to 
detect also intense emission from the Na\,{\sc i} doublet. These emission lines are 
reported from the environments of many evolved massive stars, such as YHGs 
\citep{1981ApJ...248..638L, 2013A&A...551A..69O, 2014A&A...561A..15C, 2017A&A...601A..76K, 
2022MNRAS.515.2766K, 2023Galax..11...76K}, most B[e]SGs \citep{1988ApJ...334..639M, 
2013A&A...558A..17O, 2021BAAA...62..104A}, but also LBVs in their cool (i.e. active or 
outburst) phase \citep{2013A&A...558A..17O}. Therefore, their presence does not impose any 
further constraints on the evolutionary state of an object. However, we point out that the 
reported change in the behaviour of the He\,{\sc i} lines in the optical spectrum is 
certainly worth being further monitored.

\subsection{Objects in M33}
\label{sect:M33}

\subsubsection{The LBV star Var~83}

Turning to the objects in M33, we start with J013410.93+303437.6, which is also known as M33 
Var~83 detected by \citet{1975ApJS...29..303V} as a \citet{1953ApJ...118..353H} variable. With a 
luminosity of $\log (L/L_{\odot}) = 6.3$ \citep{2014ApJ...790...48H}, J013410.93+303437.6 is 
clearly more luminous than all known B[e]SGs.
Its optical emission-line spectrum has been described by \citet{1978ApJ...219..445H} who 
found close similarity of its spectral appearance with other luminous and variable stars in M31 
and M33. From then on, this star was listed as an LBV. To our knowledge, it never underwent an 
LBV-typical eruption that would justify this clear classification, but it displays infrared 
emission, in particular at $8\,\mu$m, that offsets the star from the region occupied by classical
LBVs in the infrared colour-colour diagram \citep{2012A&A...541A.146C} and might suggest warm dust 
in its environment.

\begin{figure}
\includegraphics[width=\columnwidth]{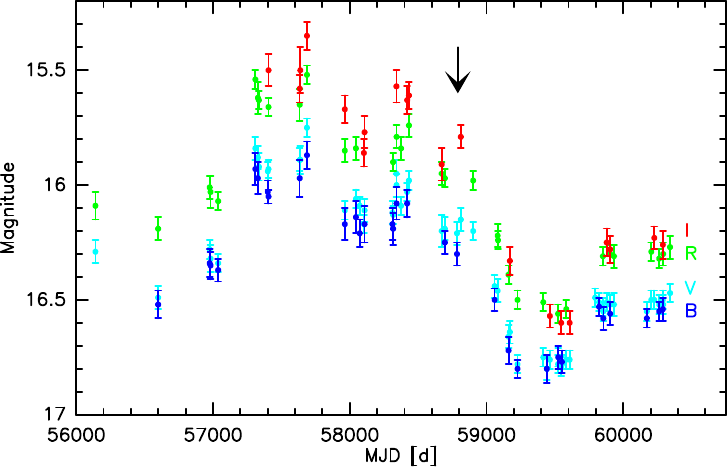}
\caption{Time series observations of J013410.93+303437.6 (Var 83) collected in $BVRI$ 
filters. The black arrow marks the date of our K-band observations.
\label{fig:Var83}}
\end{figure}

Some clues to the likely LBV nature of the star are provided by the multi-epoch $BVRI$ 
photometry, shown in Fig.~\ref{fig:Var83}. The measurements are taken from the catalogue for 
luminous stars in M33\footnote{https://uisacad5.uis.edu/~jmart5/M31M33photcat/} and spread over 
about 12 years (from July 2012 to February 2024). The light curves display irregular brightness 
variations of $0.7 - 0.8$\,mag in all bands with no or only mild color variations. And a 
$V$ band variability with an amplitude of $0.71$\,mag was already reported by 
\citet{2017AJ....154...81M}. The brightening event that started beyond MJD = 57000\,d is similar 
in amplitude to the one detected in the years 1981-1982 \citep{1996A&A...314..131S}. But in this 
recent event, it lasted for about five years. Such brightness variations, with durations of 
years to decades, are typically observed in LBVs and known as S~Dor cycles 
\citep{2020Galax...8...20W}. As mentioned before, these S~Dor cycles are usually accompanied 
by a spectroscopic change from an earlier to a later spectral type. 
However, spectroscopic observations of J013410.93+303437.6 are sparse. Interestingly, 
\citet{2017ApJ...836...64H} reported stronger P\,Cygni profiles in the He\,{\sc i} lines in the 
years 2013 and 2014, i.e. during quiescence before the brightening, compared to their spectrum 
taken in 2010. Unfortunately, the light curve does not date back to 2010, so that we cannot be 
sure whether the star was also in a brighter stage in 2010. Previous $V$-band measurements
cover the years 2000 $-$ 2003 when the magnitude showed only a mild scatter of $\pm0.1$\,mag 
around the average value of $16.44$\,mag \citep{2006MNRAS.370.1429S}.

The time of our observations was after the maximum brightness when the star was approaching 
again its quiescent state, which explains the $K$-band appearance with Pfund-line emission typical 
for a hot, luminous object. The emission from the hydrogen Pfund series that we detect is rather 
weak but comparable to the one of the similarly luminous LBV star LHA 120-S 128 in the LMC 
\citep{2013A&A...558A..17O}. Consequently, our observations align well with the likely 
classification of J013410.93+303437.6 as an LBV.

\subsubsection{A red supergiant binary with a massive companion?}

The last of our targets is J013242.26+302114.1. Its spectrum 
displays clear and pure indication of a late-type star. No emission is detected, which
complicates a refined classification of the star. The object has been previously 
classified based on its optical appearance, mainly in the blue spectral region, as a
(hot) cLBV by \citet{2007AJ....134.2474M}, 
as an iron star by \citet{2012A&A...541A.146C}, and as a B[e]SG by \citet{2017ApJ...836...64H}.  

\citet{2012A&A...541A.146C} report clear spectral variability over a 4-year period for 
J013242.26+302114.1, in particular a change in emission line ratios of the Balmer lines.
Their blue optical spectrum, acquired in September 2010, clearly indicates that the star 
is a hot, emission-line supergiant. These authors also noted photometric 
variability with an amplitude of $\Delta V \sim 0.8$\,mag with no periodicity detected 
so far \citep[see also][]{2007AJ....134.2474M}, and that the star's mid-infrared colours
suggest a wind plus photosphere rather than circumstellar dust. As 
\citet{2019Galax...7...83K} points out, the near-infrared colours place the star in the 
region of late-type stars, and she removed 
J013242.26+302114.1 from the list of B[e]SGs. Interestingly, \citet{2012ApJ...750...97D} 
classify J013242.26+302114.1 as RSG based on the star's location in colour-colour 
diagrams and the strengths of the Ca\,{\sc ii} triplet lines and the O\,{\sc i} 
$\lambda$7774 line seen in their red spectra taken in 26/27 November 
2010. Our $K$-band spectrum from January 2019 confirms the presence of a RSG. 
Therefore, we propose J013242.26+302114.1 could be a binary system consisting 
of a hot emission-line object and a RSG companion. 

To test this hypothesis, we inspect the SED of J013242.26+302114.1 to see whether a RSG could be 
present in the data. For this purpose, photometric data were collected from various sources. 
We utilize the UV catalogue of \citet{2014Ap&SS.354...97Y}, optical photometry from 
\citet{2006AJ....131.2478M}, near-infrared data from the 2MASS catalogue 
\citep{2003yCat.2246....0C, 2006AJ....131.1163S} and mid-infrared measurements from the 
{\sl Spitzer Space Telescope} \citep{2015ApJS..219...42K}. We excluded the measurements from 
WISE, because the star is in a crowded region \citep{2016AJ....152...62M}. For 
dereddening, we apply the extinction curve of \citet{1989ApJ...345..245C} with a value of 
$R_{V} = 3.2$.
To represent the stellar continuum, we use models from the grid of continuum spectra computed by 
\citet{1992IAUS..149..225K}. We find a good match of the UV and blue optical range for an 
effective temperature of $T_{\rm eff, hot} = 22\,000 - 24\,000$\,K and an extinction value of 
$A_{V} = 0.65 - 0.7$\,mag. These parameters are similar to the ones obtained by 
\citet{2017ApJ...844...40H}.

\begin{figure}
\includegraphics[width=\columnwidth]{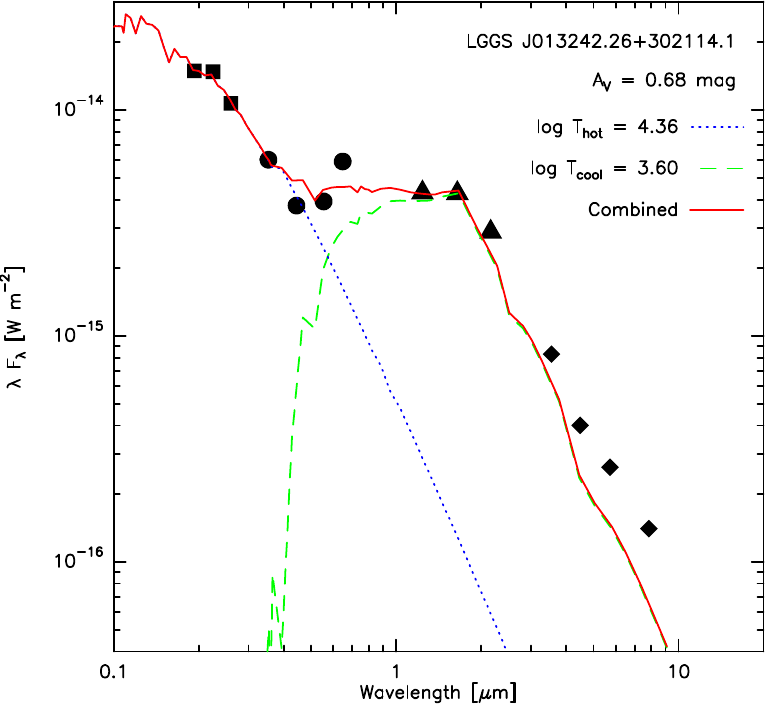}
\caption{Dereddened SED of J013242.26+302114.1, consisting of a hot object and a RSG. Photometric data are from the UV catalogue \citep[][squares]{2014Ap&SS.354...97Y}, the UBVR survey \citep[][circles]{2006AJ....131.2478M}, the 2MASS catalogue \citep[][triangles]{2003yCat.2246....0C}, and the {\sl Spitzer Space Telescope} \citep[][diamonds]{2015ApJS..219...42K}.
\label{fig:SED}}
\end{figure}

The dereddened SED of J013242.26+302114.1 is plotted in Fig.~\ref{fig:SED}. It clearly shows 
excess emission in the near- and mid-infrared, pointing towards the presence of a RSG. The model, 
which was fit to the blue part of the SED, is shown as blue dotted line.
On the cool end, the Kurucz grid is sparse. To represent the RSG star 
with parameters close to what we estimate from the $K$-band spectrum, we utilize the model for 
$T_{\rm eff} = 4000$\,K (which is the minimum temperature of the Kurucz grid) and $\log g = 1.0$. 
This model is added as the green dashed line, and the curve for the combined models is shown with 
the red solid line. Obviously, the inclusion of the RSG allows to fit the long-wavelength part of 
the SED fairly well. Moreover, the SED implies that the $K$-band region as well as the red optical 
region (longwards of $0.7\,\mu$m) is dominated by the RSG whereas the hot star dominates the blue 
spectral region. This explains why \citet{2012ApJ...750...97D} identified J013242.26+302114.1 as a 
RSG based on their spectra, whereas \citet{2014ApJ...790...48H} could see only the signature of 
the hot star in their blue spectrum\footnote{Although the red spectrum available at 
http://etacar.umn.edu/LuminousStars/M31M33/M31stars.html, which was not discussed by 
\citet{2014ApJ...790...48H}, appears to exhibit the molecular absorption bands of TiO typical for 
RSGs in the spectral range $7050-7300$\,\AA.}.

\begin{figure}
\includegraphics[width=\columnwidth]{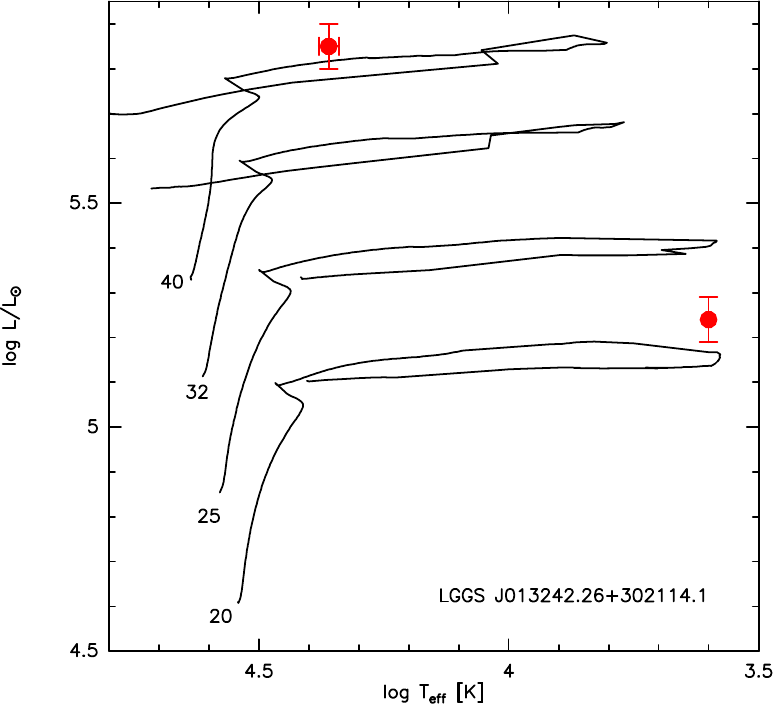}
\caption{Location of the hot and cool components of J013242.26+302114.1 in the HR diagram 
along with evolutionary tracks for rotating stars with solar metallicity 
\citep[from][]{2012A&A...537A.146E}. 
\label{fig:HRD}}
\end{figure}

From the SED fitting, and assuming a distance to M33 of $968\pm 50$\,kpc 
\citep{2009ApJ...704.1120U}, we estimate stellar luminosities of $\log L_{\rm hot}/L_{\odot} = 
5.85\pm 0.05$ and $\log L_{\rm cool}/L_{\odot} = 5.24\pm 0.05$ for the hot and cool components, 
respectively. Fig.~\ref{fig:HRD} shows the positions of the two stars in the HR diagram along with 
stellar evolutionary tracks. A binary scenario might be possible if the massive component is an 
evolved, post-RSG star. In that case, the evolutionary tracks of rotating stars in the 
initial mass range of $40-45$\,M$_{\odot}$ postulate an age of about $5.4-5.8$\,Myr 
and a current mass of $M_{\rm hot} \simeq 20-22$\,M$_{\odot}$ while the RSG age and mass 
would be on the order of $8.9-9.3$\,Myr and $M_{\rm cool} = 16.0-16.5$\,M$_{\odot}$, 
respectively, when considering evolutionary tracks for rotating stars in an initial mass 
range of $21-22$\,M$_{\odot}$. While not the same, the ages for these single-star evolution 
scenarios have at least the same order of magnitude. The deviation from exact coeval 
evolution of the two components is similar to that found, for example, for a number of RSG binary 
components in the Small Magellanic Cloud \citep{2024arXiv241218554P}.

\begin{figure}
\includegraphics[width=\columnwidth]{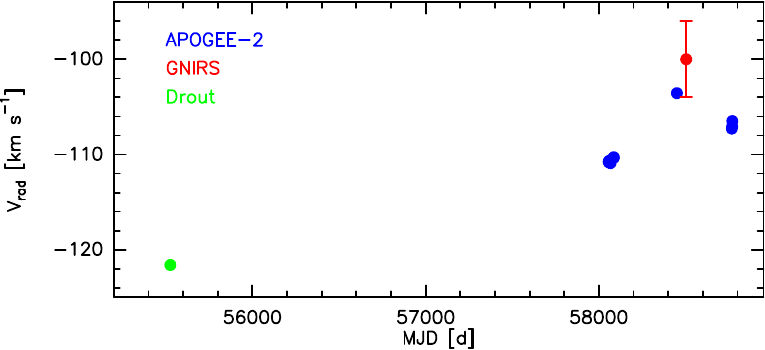}
\caption{Radial velocity curve of the RSG component in J013242.26+302114.1.
\label{fig:Vrad}}
\end{figure}

To affirm the binarity of J013242.26+302114.1, information about radial velocities is required. 
Unfortunately, no radial velocity measurements of the hot component exist, since the blue spectral 
region displays solely emission lines whose radial velocities do not necessarily reflect the 
dynamics of the hidden object\footnote{A low-resolution (69\,km\,s$^{-1}$ per pixel) optical 
spectrum covering the wavelength range $3700-9100$\,\AA \ was observed with LAMOST 
\citep{2022yCat.5156....0L}. It displays clearly the composite features of a hot emission-line 
star and a RSG. However, we observed a systematic wavelength shift between emission lines in the 
blue and the green region on the order of $15$\,km\,s$^{-1}$ that might indicate wavelength 
calibration uncertainties. Therefore, we refrain from using this spectrum for radial velocity 
measurements.}. However, in the frame of the second survey of the Apache Point Observatory 
Galactic Evolution Experiment (APOGEE-2), the RSG star has been observed multiple times between 
2017 October 30 and 2019 November 13. High-resolution ($R\sim 22\,500$) spectra were collected in 
the near-infrared H-band ($1.51 - 1.70\,\mu$m) range and radial velocity measurements were 
provided by this survey \citep{2022ApJS..259...35A}. Furthermore, \citet{2012ApJ...750...97D} had 
measured the radial velocity of the RSG in their red optical spectrum, and we measured it in our 
$K$-band spectrum. All values are listed in Table~\ref{tab:Vrad} and shown in Fig.~\ref{fig:Vrad} 
as a function of time. The RSG presents clear radial velocity variations that might support 
the binary hypothesis.

\begin{table}
	\centering
	\caption{Radial velocity measurements for the RSG component in J013242.26+302114.1.}
	\label{tab:Vrad}
\begin{tabular}{lccl}
\hline
MJD	 &  $v_{\rm rad}$   &  $\Delta v_{\rm rad}$  &  Source  \\
$ $[d]  &  [km\,s$^{-1}$]  &  [km\,s$^{-1}$]        &  \\
\hline
  55527. 	   &  $-$121.6  &   --   &   \citet{2012ApJ...750...97D}  \\
  58056.70240  &  $-$110.8  &   $<$0.2  &     APOGEE-2   \\  
  58059.69937  &  $-$110.8  &   $<$0.2  &     APOGEE-2   \\ 
  58059.87737  &  $-$110.7  &   $<$0.2  &     APOGEE-2   \\   
  58060.80436  &  $-$110.7  &   $<$0.2  &     APOGEE-2   \\  
  58067.84922  &  $-$110.9  &   $<$0.2  &     APOGEE-2   \\  
  58084.74756  &  $-$110.4  &   $<$0.2  &     APOGEE-2    \\ 
  58085.62552  &  $-$110.3  &   $<$0.2  &     APOGEE-2   \\ 
  58086.57947  &  $-$110.3  &   $<$0.2  &     APOGEE-2    \\ 
  58450.81452  &  $-$103.6  &   $<$0.2  &     APOGEE-2   \\
  58504.72323  &  $-$100.0  &      4.0  &     GNIRS  \\
  58767.71925  &  $-$107.3  &   $<$0.2  &     APOGEE-2   \\
  58769.77429  &  $-$107.1  &   $<$0.2  &     APOGEE-2   \\ 
  58770.73831  &  $-$106.5  &   $<$0.2  &     APOGEE-2   \\
\hline
\end{tabular}
\end{table}

The measurements are too sparse to reliably model the radial velocity curve, but based on 
the values from APOGEE-2 and \citet{2012ApJ...750...97D} we may claim that the semi-amplitude of 
the velocity of the RSG is at least $\sim 9$\,km\,s$^{-1}$. Moreover, we can estimate the minimum 
orbital period $P_{\rm orb, min}$, by requesting that the semi-major axis (orbital separation), 
$a$, of the binary system should be larger than the sum of the radii of the two stars. With radii 
of the hot and cool components of $R_{\rm hot} \simeq 53$\,R$_{\odot}$ and $R_{\rm cool} \simeq 
867$\,R$_{\odot}$, respectively, the minimum orbital separation is $a_{\rm min} =  R_{\rm hot} + 
R_{\rm cool} \simeq 4.28$\,AU and we obtain $P_{\rm orb, min} = 2 \pi \sqrt{a_{\rm min}^3 / 
(G  M_{\rm tot})} \simeq 532$\,d, if we use for the total mass an average value of $M_{\rm tot} 
\simeq 37$\,M$_{\odot}$. From the stellar radii and assumed current masses of the two 
components, we can infer that the orbital separation of the binary must be 
larger than the sum of the two stellar radii to satisfy the condition $a = (M_{\rm total}/M_{\rm 
hot})\,a_{\rm cool} = 1.76\,a_{\rm cool}$, or equivalently, $a_{\rm cool} = 0.57\,a$, where
$a_{\rm cool}$ is the distance of the RSG to the center of mass. Furthermore, utilizing the 
equation proposed by \citet{1983ApJ...268..368E} for determining the effective radius of the 
Roche lobe, $R_{\rm L}$, of the RSG, we find a value of $R_{\rm L,cool} = 0.82\,a$ which is 
significantly larger than $a_{\rm cool}$, meaning that the star is not filling its Roche lobe. 
This, in turn, agrees with the about coeval evolution of the two components in which binary 
interaction has not played a (significant) role in the history of this object.

If J013242.26+302114.1 is indeed a binary with a luminous hot primary and a RSG secondary 
component, this would be an extremely interesting object because it would be one of the rare cases 
of a system with an evolved primary. The RSG binaries identified so far in the Milky 
Way and galaxies of the Local Group consist of a RSG primary and a B-type main sequence 
secondary \citep[e.g.][]{2018AJ....156..225N, 2019ApJ...875..124N, 2020ApJ...900..118N, 
2020RNAAS...4...12P, 2021ApJ...908...87N, 2022MNRAS.513.5847P, 2024arXiv241218554P}. Such 
systems are to be expected from an evolutionary perspective, since companions with a lower 
mass than a B-type main-sequence star have not yet had enough time to reach the main sequence, 
while more massive companions evolve so quickly that it is rather unlikely to catch them at the 
right moment \citep[see for example the discussion in][]{2020ApJ...900..118N}. Also, many massive 
binaries merge before their primary even reaches the RSG stage, which explains why the currently 
known RSG binary fractions range from about $20\%$ in the low-metallicity environment of the SMC 
\citep{2022MNRAS.513.5847P} to about $40\%$ in the high-metallicity inner region of M33 
\citep{2021ApJ...908...87N} compared to the fraction of massive binaries on the main sequence, 
that is at least twice as high \citep[e.g.][]{2012Sci...337..444S}. Only two RSG binary systems 
are known with a different (non-main sequence) companion. These are the high-mass X-ray binary system 4U\,1954$+$31 
consisting of a RSG and a neutron star \citep{2020ApJ...904..143H} and V766\,Cen consisting of a 
RSG and a cool giant or supergiant \citep{2017A&A...606L...1W}. Hence, finding systems with 
different companion stars, in particular more massive ones, may help shedding light on the 
properties of massive binary components in such extreme evolutionary stages and their final fate.

While it is tempting to predict J013242.26+302114.1 as the first RSG binary with an 
LBV/B[e]SG primary, we wish to caution that the reported radial velocity 
variations might also have a different origin. It is well known that RSGs have semiperiodic or 
irregular light variability possibly due to the interplay of radial pulsations 
and motions of major convection 
cells, as was proposed for Betelgeuse \citep{2008AJ....135.1450G}. Periods were found to range 
from hundreds to thousands of days \citep[e.g.][]{2006MNRAS.372.1721K, 2019MNRAS.487.4832C} and 
radial velocity variations of up to $\sim 10$\,km\,s$^{-1}$ have been detected in numerous RSGs 
\citep[e.g.][]{2007A&A...469..671J, 2018ApJ...859...73S} and were predicted by 1D pulsation test 
models \citep{2015A&A...575A..50A}. The radial pulsations in RSGs follow a period-luminosity relation according to which J013242.26+302114.1, having a $K$-band absolute magnitude of $M_{K} \sim -11$\,mag, is expected to experience radial pulsations with a period of $500-1000$\,d 
\citep{2006MNRAS.372.1721K, 2019MNRAS.487.4832C}. Due to this ambiguity,
follow-up observations of J013242.26+302114.1 are urgently needed to complete the 
radial velocity curve, confirm that the hot, massive star is indeed the companion and not 
just a line-of-sight projection, and if confirmed, constrain the orbital parameters and stellar 
masses that may shed light on the evolutionary history of this interesting object. 
For this, UV spectroscopy might help to characterize the hot component in
J013242.26+302114.1 \citep{2024arXiv241218554P} and to measure its radial velocity.

\section{Conclusions}

%The last numbered section should briefly summarise what has been done, and describe
%the final conclusions which the authors draw from their work.

We have carried out a $K$-band spectroscopic survey of six evolved massive stars in M31 and 
M33. Five of them had previously uncertain or controversial classification and one is an 
LBV. We have detected dense and warm molecular gas rings around two objects in M31 
(J004320.97+414039.6 and J004621.08+421308.2) that we classify as B[e]SGs. Modelling the CO band 
emission revealed the physical properties (temperature, column density, gas dynamics) of the 
line-forming regions in these rings together with information about the surface enrichment in 
$^{13}$C of the star at the time when the material, forming these rings, was ejected. Based 
on the measured high enrichment for J004320.97+414039.6, we conclude that the star is most likely 
in post-RSG evolution. We have detected a dense ionized wind from the M31 star J004415.00+420156.2 
that we also consider as B[e]SG and a weaker one from the LBV star J013410.93+303437.6 
(Var~83) in M33. The latter object just returned from an S~Dor cycle at the time of our 
observations, explaining possibly the weaker wind observed. For the M31 object J004229.87+410551.8 
with the featureless near-infrared spectrum, we investigated whether it could be a B[e]SG or an 
LBV. The lack of $K$-band features together with a constant brightness make it difficult to 
unambiguously determine the true nature of the object. However, the inspection of its previous 
photometry suggests that the star may be an LBV undergoing an S~Dor cycle. In addition, we 
discovered that J013242.26+302114.1 in M33 is most likely a binary system consisting of a hot 
B-type emission-line object, which could be a post-RSG in an LBV or a B[e]SG phase, and a 
RSG companion. If confirmed, it would be the first RSG binary with a massive LBV/B[e]SG primary. 
Follow-up measurements are therefore urgently needed to complete the so far sparsely populated 
radial velocity curve and to validate the binary over the pulsation scenario. Once the orbital 
parameters are constrained, they could provide insight into the evolutionary history and possible 
fate of this interesting object.

The presented data and analyses reinforce that near-infrared spectra add significant
information that helps classifying evolved massive stars, characterizing circumstellar 
environments, as well as identifying late-type companions, which can be missed out from the 
optical observations.

\section*{Acknowledgements}

We thank the reviewer, Lee Patrick, for valuable comments and suggestions on the manuscript.

This paper is based on observations obtained at the international Gemini Observatory, a program of NSF NOIRLab, which is managed by the Association of Universities for Research in Astronomy (AURA) under a cooperative agreement with the U.S. National Science Foundation on behalf of the Gemini Observatory partnership: the U.S. National Science Foundation (United States), National Research Council (Canada), Agencia Nacional de Investigaci\'{o}n y Desarrollo (Chile), Ministerio de Ciencia, Tecnolog\'{i}a e Innovaci\'{o}n (Argentina), Minist\'{e}rio da Ci\^{e}ncia, Tecnologia e Inova\c{c}\~{o}es (Brazil), and Korea Astronomy and Space Science Institute (Republic of Korea) under program IDs GN-2018B-Q-301, GN-2019B-Q-201, and GN-2020B-Q-230.

This research made use of the NASA Astrophysics Data System (ADS), of the SIMBAD database, 
operated at CDS, Strasbourg, France, of data products from the Two Micron All Sky Survey, which is 
a joint project of the University of Massachusetts and the Infrared Processing and Analysis 
Center/California Institute of Technology, funded by the National Aeronautics and Space 
Administration and the National Science Foundation, and of data from the Zwicky Transient Facility 
(ZTF). The ZTF is supported by the National Science Foundation under Grants No. AST-1440341 and 
AST-2034437 and a collaboration including current partners Caltech, IPAC, the Weizmann Institute 
of Science, the Oskar Klein Center at Stockholm University, the University of Maryland, Deutsches 
Elektronen-Synchrotron and Humboldt University, the TANGO Consortium of Taiwan, the University of 
Wisconsin at Milwaukee, Trinity College Dublin, Lawrence Livermore National Laboratories, IN2P3, 
University of Warwick, Ruhr University Bochum, Northwestern University and former partners the 
University of Washington, Los Alamos National Laboratories, and Lawrence Berkeley National 
Laboratories. Operations are conducted by COO, IPAC, and UW.

The Astronomical Institute of the Czech Academy of Sciences is supported by the project 
RVO:67985815. MLA and AFT acknowledge financial support from CONICET (PIP 1337) and the 
Universidad Nacional de La Plata (Programa de Incentivos 11/G192), Argentina. MBF acknowledges 
financial support from the National Council for Scientific and Technological Development - 
CNPq - Brazil (grant number: 307711/2022-6). This project 
has received funding from the European Union's Framework Programme for Research and 
Innovation Horizon 2020 (2014-2020) under the Marie Sk\l{}odowska-Curie Grant Agreement 
No. 823734 (POEMS) and from the project No. 101183150 (OCEANS).

%%%%%%%%%%%%%%%%%%%%%%%%%%%%%%%%%%%%%%%%%%%%%%%%%%
\section*{Data Availability}

The observed GNIRS $K$-band spectra can be retrieved from the publicly available Gemini 
Observatory Archive (https://archive.gemini.edu). The processed spectra will be shared on a 
reasonable request to the corresponding author.

%The inclusion of a Data Availability Statement is a requirement for articles published in MNRAS. Data Availability Statements provide a standardised format for readers to understand the availability of data underlying the research results described in the article. The statement may refer to original data generated in the course of the study or to third-party data analysed in the article. The statement should describe and provide means of access, where possible, by linking to the data or providing the required accession numbers for the relevant databases or DOIs.



\bibliographystyle{mnras}
\bibliography{M31_M33} % if your bibtex file is called example.bib

\begin{thebibliography}{}
\makeatletter
\relax
\def\mn@urlcharsother{\let\do\@makeother \do\$\do\&\do\#\do\^\do\_\do\%\do\~}
\def\mn@doi{\begingroup\mn@urlcharsother \@ifnextchar [ {\mn@doi@}
  {\mn@doi@[]}}
\def\mn@doi@[#1]#2{\def\@tempa{#1}\ifx\@tempa\@empty \href
  {http://dx.doi.org/#2} {doi:#2}\else \href {http://dx.doi.org/#2} {#1}\fi
  \endgroup}
\def\mn@eprint#1#2{\mn@eprint@#1:#2::\@nil}
\def\mn@eprint@arXiv#1{\href {http://arxiv.org/abs/#1} {{\tt arXiv:#1}}}
\def\mn@eprint@dblp#1{\href {http://dblp.uni-trier.de/rec/bibtex/#1.xml}
  {dblp:#1}}
\def\mn@eprint@#1:#2:#3:#4\@nil{\def\@tempa {#1}\def\@tempb {#2}\def\@tempc
  {#3}\ifx \@tempc \@empty \let \@tempc \@tempb \let \@tempb \@tempa \fi \ifx
  \@tempb \@empty \def\@tempb {arXiv}\fi \@ifundefined
  {mn@eprint@\@tempb}{\@tempb:\@tempc}{\expandafter \expandafter \csname
  mn@eprint@\@tempb\endcsname \expandafter{\@tempc}}}

\bibitem[\protect\citeauthoryear{{Abdurro'uf} et~al.,}{{Abdurro'uf}
  et~al.}{2022}]{2022ApJS..259...35A}
{Abdurro'uf} et~al., 2022, \mn@doi [\apjs] {10.3847/1538-4365/ac4414}, \href
  {https://ui.adsabs.harvard.edu/abs/2022ApJS..259...35A} {259, 35}

\bibitem[\protect\citeauthoryear{{Aret}, {Kraus}, {Muratore}  \& {Borges
  Fernandes}}{{Aret} et~al.}{2012}]{2012MNRAS.423..284A}
{Aret} A.,  {Kraus} M.,  {Muratore} M.~F.,   {Borges Fernandes} M.,  2012,
  \mn@doi [\mnras] {10.1111/j.1365-2966.2012.20871.x}, \href
  {https://ui.adsabs.harvard.edu/abs/2012MNRAS.423..284A} {423, 284}

\bibitem[\protect\citeauthoryear{{Arias}, {Vallverd{\'u}}, {Torres}  \&
  {Kraus}}{{Arias} et~al.}{2021}]{2021BAAA...62..104A}
{Arias} M.~L.,  {Vallverd{\'u}} R.,  {Torres} A.~F.,   {Kraus} M.,  2021,
  Boletin de la Asociacion Argentina de Astronomia La Plata Argentina, \href
  {https://ui.adsabs.harvard.edu/abs/2021BAAA...62..104A} {62, 104}

\bibitem[\protect\citeauthoryear{{Arroyo-Torres} et~al.,}{{Arroyo-Torres}
  et~al.}{2015}]{2015A&A...575A..50A}
{Arroyo-Torres} B.,  et~al., 2015, \mn@doi [\aap]
  {10.1051/0004-6361/201425212}, \href
  {https://ui.adsabs.harvard.edu/abs/2015A&A...575A..50A} {575, A50}

\bibitem[\protect\citeauthoryear{{Bellm} et~al.,}{{Bellm}
  et~al.}{2019}]{2019PASP..131f8003B}
{Bellm} E.~C.,  et~al., 2019, \mn@doi [\pasp] {10.1088/1538-3873/ab0c2a}, \href
  {https://ui.adsabs.harvard.edu/abs/2019PASP..131f8003B} {131, 068003}

\bibitem[\protect\citeauthoryear{{Bonanos} et~al.,}{{Bonanos}
  et~al.}{2024}]{2024A&A...686A..77B}
{Bonanos} A.~Z.,  et~al., 2024, \mn@doi [\aap] {10.1051/0004-6361/202348527},
  \href {https://ui.adsabs.harvard.edu/abs/2024A&A...686A..77B} {686, A77}

\bibitem[\protect\citeauthoryear{{Cardelli}, {Clayton}  \& {Mathis}}{{Cardelli}
  et~al.}{1989}]{1989ApJ...345..245C}
{Cardelli} J.~A.,  {Clayton} G.~C.,   {Mathis} J.~S.,  1989, \mn@doi [\apj]
  {10.1086/167900}, \href
  {https://ui.adsabs.harvard.edu/abs/1989ApJ...345..245C} {345, 245}

\bibitem[\protect\citeauthoryear{{Carr}}{{Carr}}{1995}]{1995Ap&SS.224...25C}
{Carr} J.~S.,  1995, \mn@doi [\apss] {10.1007/BF00667816}, \href
  {https://ui.adsabs.harvard.edu/abs/1995Ap&SS.224...25C} {224, 25}

\bibitem[\protect\citeauthoryear{{Chatys}, {Bedding}, {Murphy}, {Kiss}, {Dobie}
   \& {Grindlay}}{{Chatys} et~al.}{2019}]{2019MNRAS.487.4832C}
{Chatys} F.~W.,  {Bedding} T.~R.,  {Murphy} S.~J.,  {Kiss} L.~L.,  {Dobie} D.,
   {Grindlay} J.~E.,  2019, \mn@doi [\mnras] {10.1093/mnras/stz1584}, \href
  {https://ui.adsabs.harvard.edu/abs/2019MNRAS.487.4832C} {487, 4832}

\bibitem[\protect\citeauthoryear{{Cidale} et~al.,}{{Cidale}
  et~al.}{2012}]{2012A&A...548A..72C}
{Cidale} L.~S.,  et~al., 2012, \mn@doi [\aap] {10.1051/0004-6361/201220120},
  \href {https://ui.adsabs.harvard.edu/abs/2012A&A...548A..72C} {548, A72}

\bibitem[\protect\citeauthoryear{{Clark}, {Castro}, {Garcia}, {Herrero},
  {Najarro}, {Negueruela}, {Ritchie}  \& {Smith}}{{Clark}
  et~al.}{2012}]{2012A&A...541A.146C}
{Clark} J.~S.,  {Castro} N.,  {Garcia} M.,  {Herrero} A.,  {Najarro} F.,
  {Negueruela} I.,  {Ritchie} B.~W.,   {Smith} K.~T.,  2012, \mn@doi [\aap]
  {10.1051/0004-6361/201118440}, \href
  {https://ui.adsabs.harvard.edu/abs/2012A&A...541A.146C} {541, A146}

\bibitem[\protect\citeauthoryear{{Clark}, {Negueruela}  \&
  {Gonz{\'a}lez-Fern{\'a}ndez}}{{Clark} et~al.}{2014}]{2014A&A...561A..15C}
{Clark} J.~S.,  {Negueruela} I.,   {Gonz{\'a}lez-Fern{\'a}ndez} C.,  2014,
  \mn@doi [\aap] {10.1051/0004-6361/201322772}, \href
  {https://ui.adsabs.harvard.edu/abs/2014A&A...561A..15C} {561, A15}

\bibitem[\protect\citeauthoryear{{Cochetti}, {Kraus}, {Arias}, {Cidale},
  {Eenm{\"a}e}, {Liimets}, {Torres}  \& {Djupvik}}{{Cochetti}
  et~al.}{2020}]{2020AJ....160..166C}
{Cochetti} Y.~R.,  {Kraus} M.,  {Arias} M.~L.,  {Cidale} L.~S.,  {Eenm{\"a}e}
  T.,  {Liimets} T.,  {Torres} A.~F.,   {Djupvik} A.~A.,  2020, \mn@doi [\aj]
  {10.3847/1538-3881/abae62}, \href
  {https://ui.adsabs.harvard.edu/abs/2020AJ....160..166C} {160, 166}

\bibitem[\protect\citeauthoryear{{Corral}}{{Corral}}{1996}]{1996AJ....112.1450C}
{Corral} L.~J.,  1996, \mn@doi [\aj] {10.1086/118113}, \href
  {https://ui.adsabs.harvard.edu/abs/1996AJ....112.1450C} {112, 1450}

\bibitem[\protect\citeauthoryear{{Cutri} et~al.,}{{Cutri}
  et~al.}{2003}]{2003yCat.2246....0C}
{Cutri} R.~M.,  et~al., 2003, {VizieR Online Data Catalog: 2MASS All-Sky
  Catalog of Point Sources (Cutri+ 2003)}, VizieR On-line Data Catalog: II/246.
  Originally published in: 2003yCat.2246....0C

\bibitem[\protect\citeauthoryear{{Dorn-Wallenstein}, {Neugent}  \&
  {Levesque}}{{Dorn-Wallenstein} et~al.}{2023}]{2023ApJ...959..102D}
{Dorn-Wallenstein} T.~Z.,  {Neugent} K.~F.,   {Levesque} E.~M.,  2023, \mn@doi
  [\apj] {10.3847/1538-4357/ad0725}, \href
  {https://ui.adsabs.harvard.edu/abs/2023ApJ...959..102D} {959, 102}

\bibitem[\protect\citeauthoryear{{Drout}, {Massey}  \& {Meynet}}{{Drout}
  et~al.}{2012}]{2012ApJ...750...97D}
{Drout} M.~R.,  {Massey} P.,   {Meynet} G.,  2012, \mn@doi [\apj]
  {10.1088/0004-637X/750/2/97}, \href
  {https://ui.adsabs.harvard.edu/abs/2012ApJ...750...97D} {750, 97}

\bibitem[\protect\citeauthoryear{{Eggleton}}{{Eggleton}}{1983}]{1983ApJ...268..368E}
{Eggleton} P.~P.,  1983, \mn@doi [\apj] {10.1086/160960}, \href
  {https://ui.adsabs.harvard.edu/abs/1983ApJ...268..368E} {268, 368}

\bibitem[\protect\citeauthoryear{{Ekstr{\"o}m} et~al.,}{{Ekstr{\"o}m}
  et~al.}{2012}]{2012A&A...537A.146E}
{Ekstr{\"o}m} S.,  et~al., 2012, \mn@doi [\aap] {10.1051/0004-6361/201117751},
  \href {https://ui.adsabs.harvard.edu/abs/2012A&A...537A.146E} {537, A146}

\bibitem[\protect\citeauthoryear{{Eldridge} \& {Stanway}}{{Eldridge} \&
  {Stanway}}{2022}]{2022ARA&A..60..455E}
{Eldridge} J.~J.,  {Stanway} E.~R.,  2022, \mn@doi [\araa]
  {10.1146/annurev-astro-052920-100646}, \href
  {https://ui.adsabs.harvard.edu/abs/2022ARA&A..60..455E} {60, 455}

\bibitem[\protect\citeauthoryear{{Elias}, {Rodgers}, {Joyce}, {Lazo},
  {Doppmann}, {Winge}  \& {Rodr{\'\i}guez-Ardila}}{{Elias}
  et~al.}{2006a}]{2006SPIE.6269E..14E}
{Elias} J.~H.,  {Rodgers} B.,  {Joyce} R.~R.,  {Lazo} M.,  {Doppmann} G.,
  {Winge} C.,   {Rodr{\'\i}guez-Ardila} A.,  2006a, in {McLean} I.~S.,  {Iye}
  M.,  eds,  Society of Photo-Optical Instrumentation Engineers (SPIE)
  Conference Series Vol. 6269, Ground-based and Airborne Instrumentation for
  Astronomy. p. 626914, \mn@doi{10.1117/12.671765}

\bibitem[\protect\citeauthoryear{{Elias}, {Joyce}, {Liang}, {Muller}, {Hileman}
   \& {George}}{{Elias} et~al.}{2006b}]{2006SPIE.6269E..4CE}
{Elias} J.~H.,  {Joyce} R.~R.,  {Liang} M.,  {Muller} G.~P.,  {Hileman} E.~A.,
   {George} J.~R.,  2006b, in {McLean} I.~S.,  {Iye} M.,  eds,  Society of
  Photo-Optical Instrumentation Engineers (SPIE) Conference Series Vol. 6269,
  Ground-based and Airborne Instrumentation for Astronomy. p. 62694C,
  \mn@doi{10.1117/12.671817}

\bibitem[\protect\citeauthoryear{{Fabrika} \& {Sholukhova}}{{Fabrika} \&
  {Sholukhova}}{1999}]{1999A&AS..140..309F}
{Fabrika} S.,  {Sholukhova} O.,  1999, \mn@doi [\aaps] {10.1051/aas:1999425},
  \href {https://ui.adsabs.harvard.edu/abs/1999A&AS..140..309F} {140, 309}

\bibitem[\protect\citeauthoryear{{Fabrika}, {Sholukhova}, {Becker},
  {Afanasiev}, {Roth}  \& {Sanchez}}{{Fabrika}
  et~al.}{2005}]{2005A&A...437..217F}
{Fabrika} S.,  {Sholukhova} O.,  {Becker} T.,  {Afanasiev} V.,  {Roth} M.,
  {Sanchez} S.~F.,  2005, \mn@doi [\aap] {10.1051/0004-6361:20035824}, \href
  {https://ui.adsabs.harvard.edu/abs/2005A&A...437..217F} {437, 217}

\bibitem[\protect\citeauthoryear{{Georgy}, {Saio}  \& {Meynet}}{{Georgy}
  et~al.}{2014}]{2014MNRAS.439L...6G}
{Georgy} C.,  {Saio} H.,   {Meynet} G.,  2014, \mn@doi [\mnras]
  {10.1093/mnrasl/slt165}, \href
  {https://ui.adsabs.harvard.edu/abs/2014MNRAS.439L...6G} {439, L6}

\bibitem[\protect\citeauthoryear{{Gordon}, {Humphreys}  \& {Jones}}{{Gordon}
  et~al.}{2016}]{2016ApJ...825...50G}
{Gordon} M.~S.,  {Humphreys} R.~M.,   {Jones} T.~J.,  2016, \mn@doi [\apj]
  {10.3847/0004-637X/825/1/50}, \href
  {https://ui.adsabs.harvard.edu/abs/2016ApJ...825...50G} {825, 50}

\bibitem[\protect\citeauthoryear{{Gray}}{{Gray}}{2008}]{2008AJ....135.1450G}
{Gray} D.~F.,  2008, \mn@doi [\aj] {10.1088/0004-6256/135/4/1450}, \href
  {https://ui.adsabs.harvard.edu/abs/2008AJ....135.1450G} {135, 1450}

\bibitem[\protect\citeauthoryear{{Groh}, {Meynet}, {Ekstr{\"o}m}  \&
  {Georgy}}{{Groh} et~al.}{2014}]{2014A&A...564A..30G}
{Groh} J.~H.,  {Meynet} G.,  {Ekstr{\"o}m} S.,   {Georgy} C.,  2014, \mn@doi
  [\aap] {10.1051/0004-6361/201322573}, \href
  {https://ui.adsabs.harvard.edu/abs/2014A&A...564A..30G} {564, A30}

\bibitem[\protect\citeauthoryear{{Hanson}, {Conti}  \& {Rieke}}{{Hanson}
  et~al.}{1996}]{1996ApJS..107..281H}
{Hanson} M.~M.,  {Conti} P.~S.,   {Rieke} M.~J.,  1996, \mn@doi [\apjs]
  {10.1086/192366}, \href
  {https://ui.adsabs.harvard.edu/abs/1996ApJS..107..281H} {107, 281}

\bibitem[\protect\citeauthoryear{{Heger} \& {Langer}}{{Heger} \&
  {Langer}}{2000}]{2000ApJ...544.1016H}
{Heger} A.,  {Langer} N.,  2000, \mn@doi [\apj] {10.1086/317239}, \href
  {https://ui.adsabs.harvard.edu/abs/2000ApJ...544.1016H} {544, 1016}

\bibitem[\protect\citeauthoryear{{Heger}, {Langer}  \& {Woosley}}{{Heger}
  et~al.}{2000}]{2000ApJ...528..368H}
{Heger} A.,  {Langer} N.,   {Woosley} S.~E.,  2000, \mn@doi [\apj]
  {10.1086/308158}, \href
  {https://ui.adsabs.harvard.edu/abs/2000ApJ...528..368H} {528, 368}

\bibitem[\protect\citeauthoryear{{Hinkle}, {Lebzelter}, {Fekel}, {Straniero},
  {Joyce}, {Prato}, {Karnath}  \& {Habel}}{{Hinkle}
  et~al.}{2020}]{2020ApJ...904..143H}
{Hinkle} K.~H.,  {Lebzelter} T.,  {Fekel} F.~C.,  {Straniero} O.,  {Joyce}
  R.~R.,  {Prato} L.,  {Karnath} N.,   {Habel} N.,  2020, \mn@doi [\apj]
  {10.3847/1538-4357/abbe01}, \href
  {https://ui.adsabs.harvard.edu/abs/2020ApJ...904..143H} {904, 143}

\bibitem[\protect\citeauthoryear{{Hubble} \& {Sandage}}{{Hubble} \&
  {Sandage}}{1953}]{1953ApJ...118..353H}
{Hubble} E.,  {Sandage} A.,  1953, \mn@doi [\apj] {10.1086/145764}, \href
  {https://ui.adsabs.harvard.edu/abs/1953ApJ...118..353H} {118, 353}

\bibitem[\protect\citeauthoryear{{Humphreys}}{{Humphreys}}{1978}]{1978ApJ...219..445H}
{Humphreys} R.~M.,  1978, \mn@doi [\apj] {10.1086/155797}, \href
  {https://ui.adsabs.harvard.edu/abs/1978ApJ...219..445H} {219, 445}

\bibitem[\protect\citeauthoryear{{Humphreys} \& {Davidson}}{{Humphreys} \&
  {Davidson}}{1994}]{1994PASP..106.1025H}
{Humphreys} R.~M.,  {Davidson} K.,  1994, \mn@doi [\pasp] {10.1086/133478},
  \href {https://ui.adsabs.harvard.edu/abs/1994PASP..106.1025H} {106, 1025}

\bibitem[\protect\citeauthoryear{{Humphreys}, {Massey}  \&
  {Freedman}}{{Humphreys} et~al.}{1990}]{1990AJ.....99...84H}
{Humphreys} R.~M.,  {Massey} P.,   {Freedman} W.~L.,  1990, \mn@doi [\aj]
  {10.1086/115315}, \href
  {https://ui.adsabs.harvard.edu/abs/1990AJ.....99...84H} {99, 84}

\bibitem[\protect\citeauthoryear{{Humphreys}, {Davidson}, {Grammer},
  {Kneeland}, {Martin}, {Weis}  \& {Burggraf}}{{Humphreys}
  et~al.}{2013}]{2013ApJ...773...46H}
{Humphreys} R.~M.,  {Davidson} K.,  {Grammer} S.,  {Kneeland} N.,  {Martin}
  J.~C.,  {Weis} K.,   {Burggraf} B.,  2013, \mn@doi [\apj]
  {10.1088/0004-637X/773/1/46}, \href
  {https://ui.adsabs.harvard.edu/abs/2013ApJ...773...46H} {773, 46}

\bibitem[\protect\citeauthoryear{{Humphreys}, {Weis}, {Davidson}, {Bomans}  \&
  {Burggraf}}{{Humphreys} et~al.}{2014}]{2014ApJ...790...48H}
{Humphreys} R.~M.,  {Weis} K.,  {Davidson} K.,  {Bomans} D.~J.,   {Burggraf}
  B.,  2014, \mn@doi [\apj] {10.1088/0004-637X/790/1/48}, \href
  {https://ui.adsabs.harvard.edu/abs/2014ApJ...790...48H} {790, 48}

\bibitem[\protect\citeauthoryear{{Humphreys}, {Gordon}, {Martin}, {Weis}  \&
  {Hahn}}{{Humphreys} et~al.}{2017a}]{2017ApJ...836...64H}
{Humphreys} R.~M.,  {Gordon} M.~S.,  {Martin} J.~C.,  {Weis} K.,   {Hahn} D.,
  2017a, \mn@doi [\apj] {10.3847/1538-4357/aa582e}, \href
  {https://ui.adsabs.harvard.edu/abs/2017ApJ...836...64H} {836, 64}

\bibitem[\protect\citeauthoryear{{Humphreys}, {Davidson}, {Hahn}, {Martin}  \&
  {Weis}}{{Humphreys} et~al.}{2017b}]{2017ApJ...844...40H}
{Humphreys} R.~M.,  {Davidson} K.,  {Hahn} D.,  {Martin} J.~C.,   {Weis} K.,
  2017b, \mn@doi [\apj] {10.3847/1538-4357/aa7cef}, \href
  {https://ui.adsabs.harvard.edu/abs/2017ApJ...844...40H} {844, 40}

\bibitem[\protect\citeauthoryear{{Humphreys}, {Stangl}, {Gordon}, {Davidson}
  \& {Grammer}}{{Humphreys} et~al.}{2019}]{2019AJ....157...22H}
{Humphreys} R.~M.,  {Stangl} S.,  {Gordon} M.~S.,  {Davidson} K.,   {Grammer}
  S.~H.,  2019, \mn@doi [\aj] {10.3847/1538-3881/aaf1ac}, \href
  {https://ui.adsabs.harvard.edu/abs/2019AJ....157...22H} {157, 22}

\bibitem[\protect\citeauthoryear{{Josselin} \& {Plez}}{{Josselin} \&
  {Plez}}{2007}]{2007A&A...469..671J}
{Josselin} E.,  {Plez} B.,  2007, \mn@doi [\aap] {10.1051/0004-6361:20066353},
  \href {https://ui.adsabs.harvard.edu/abs/2007A&A...469..671J} {469, 671}

\bibitem[\protect\citeauthoryear{{Kang}, {Rey}, {Bianchi}, {Lee}, {Kim}  \&
  {Sohn}}{{Kang} et~al.}{2012}]{2012ApJS..199...37K}
{Kang} Y.,  {Rey} S.-C.,  {Bianchi} L.,  {Lee} K.,  {Kim} Y.,   {Sohn} S.~T.,
  2012, \mn@doi [\apjs] {10.1088/0067-0049/199/2/37}, \href
  {https://ui.adsabs.harvard.edu/abs/2012ApJS..199...37K} {199, 37}

\bibitem[\protect\citeauthoryear{{Khan}, {Stanek}, {Kochanek}  \&
  {Sonneborn}}{{Khan} et~al.}{2015}]{2015ApJS..219...42K}
{Khan} R.,  {Stanek} K.~Z.,  {Kochanek} C.~S.,   {Sonneborn} G.,  2015, \mn@doi
  [\apjs] {10.1088/0067-0049/219/2/42}, \href
  {https://ui.adsabs.harvard.edu/abs/2015ApJS..219...42K} {219, 42}

\bibitem[\protect\citeauthoryear{{King}, {Walterbos}  \& {Braun}}{{King}
  et~al.}{1998}]{1998ApJ...507..210K}
{King} N.~L.,  {Walterbos} R.~A.~M.,   {Braun} R.,  1998, \mn@doi [\apj]
  {10.1086/306296}, \href
  {https://ui.adsabs.harvard.edu/abs/1998ApJ...507..210K} {507, 210}

\bibitem[\protect\citeauthoryear{{Kiss}, {Szab{\'o}}  \& {Bedding}}{{Kiss}
  et~al.}{2006}]{2006MNRAS.372.1721K}
{Kiss} L.~L.,  {Szab{\'o}} G.~M.,   {Bedding} T.~R.,  2006, \mn@doi [\mnras]
  {10.1111/j.1365-2966.2006.10973.x}, \href
  {https://ui.adsabs.harvard.edu/abs/2006MNRAS.372.1721K} {372, 1721}

\bibitem[\protect\citeauthoryear{{Koumpia}, {Oudmaijer}, {de Wit},
  {M{\'e}rand}, {Black}  \& {Ababakr}}{{Koumpia}
  et~al.}{2022}]{2022MNRAS.515.2766K}
{Koumpia} E.,  {Oudmaijer} R.~D.,  {de Wit} W.~J.,  {M{\'e}rand} A.,  {Black}
  J.~H.,   {Ababakr} K.~M.,  2022, \mn@doi [\mnras] {10.1093/mnras/stac1998},
  \href {https://ui.adsabs.harvard.edu/abs/2022MNRAS.515.2766K} {515, 2766}

\bibitem[\protect\citeauthoryear{{Kourniotis}, {Bonanos}, {Yuan}, {Macri},
  {Garcia-Alvarez}  \& {Lee}}{{Kourniotis} et~al.}{2017}]{2017A&A...601A..76K}
{Kourniotis} M.,  {Bonanos} A.~Z.,  {Yuan} W.,  {Macri} L.~M.,
  {Garcia-Alvarez} D.,   {Lee} C.~H.,  2017, \mn@doi [\aap]
  {10.1051/0004-6361/201629146}, \href
  {https://ui.adsabs.harvard.edu/abs/2017A&A...601A..76K} {601, A76}

\bibitem[\protect\citeauthoryear{{Kourniotis}, {Kraus}, {Arias}, {Cidale}  \&
  {Torres}}{{Kourniotis} et~al.}{2018}]{2018MNRAS.480.3706K}
{Kourniotis} M.,  {Kraus} M.,  {Arias} M.~L.,  {Cidale} L.,   {Torres} A.~F.,
  2018, \mn@doi [\mnras] {10.1093/mnras/sty2087}, \href
  {https://ui.adsabs.harvard.edu/abs/2018MNRAS.480.3706K} {480, 3706}

\bibitem[\protect\citeauthoryear{{Kourniotis}, {Kraus}, {Arias}  \&
  {Cidale}}{{Kourniotis} et~al.}{2025}]{2025MNRAS.540L..28K}
{Kourniotis} M.,  {Kraus} M.,  {Arias} M.~L.,   {Cidale} L.~S.,  2025, \mn@doi
  [\mnras] {10.1093/mnrasl/slaf028}, \href
  {https://ui.adsabs.harvard.edu/abs/2025MNRAS.540L..28K} {540, L28}

\bibitem[\protect\citeauthoryear{{Kraus}}{{Kraus}}{2009}]{2009A&A...494..253K}
{Kraus} M.,  2009, \mn@doi [\aap] {10.1051/0004-6361:200811020}, \href
  {https://ui.adsabs.harvard.edu/abs/2009A&A...494..253K} {494, 253}

\bibitem[\protect\citeauthoryear{{Kraus}}{{Kraus}}{2019}]{2019Galax...7...83K}
{Kraus} M.,  2019, \mn@doi [Galaxies] {10.3390/galaxies7040083}, \href
  {https://ui.adsabs.harvard.edu/abs/2019Galax...7...83K} {7, 83}

\bibitem[\protect\citeauthoryear{{Kraus}, {Kr{\"u}gel}, {Thum}  \&
  {Geballe}}{{Kraus} et~al.}{2000}]{2000A&A...362..158K}
{Kraus} M.,  {Kr{\"u}gel} E.,  {Thum} C.,   {Geballe} T.~R.,  2000, \mn@doi
  [\aap] {10.48550/arXiv.astro-ph/0008213}, \href
  {https://ui.adsabs.harvard.edu/abs/2000A&A...362..158K} {362, 158}

\bibitem[\protect\citeauthoryear{{Kraus}, {Borges Fernandes}  \& {de
  Ara{\'u}jo}}{{Kraus} et~al.}{2010}]{2010A&A...517A..30K}
{Kraus} M.,  {Borges Fernandes} M.,   {de Ara{\'u}jo} F.~X.,  2010, \mn@doi
  [\aap] {10.1051/0004-6361/200913964}, \href
  {https://ui.adsabs.harvard.edu/abs/2010A&A...517A..30K} {517, A30}

\bibitem[\protect\citeauthoryear{{Kraus}, {Oksala}, {Nickeler}, {Muratore},
  {Borges Fernandes}, {Aret}, {Cidale}  \& {de Wit}}{{Kraus}
  et~al.}{2013}]{2013A&A...549A..28K}
{Kraus} M.,  {Oksala} M.~E.,  {Nickeler} D.~H.,  {Muratore} M.~F.,  {Borges
  Fernandes} M.,  {Aret} A.,  {Cidale} L.~S.,   {de Wit} W.~J.,  2013, \mn@doi
  [\aap] {10.1051/0004-6361/201220442}, \href
  {https://ui.adsabs.harvard.edu/abs/2013A&A...549A..28K} {549, A28}

\bibitem[\protect\citeauthoryear{{Kraus}, {Cidale}, {Arias}, {Oksala}  \&
  {Borges Fernandes}}{{Kraus} et~al.}{2014}]{2014ApJ...780L..10K}
{Kraus} M.,  {Cidale} L.~S.,  {Arias} M.~L.,  {Oksala} M.~E.,   {Borges
  Fernandes} M.,  2014, \mn@doi [\apjl] {10.1088/2041-8205/780/1/L10}, \href
  {http://adsabs.harvard.edu/abs/2014ApJ...780L..10K} {780, L10}

\bibitem[\protect\citeauthoryear{{Kraus} et~al.,}{{Kraus}
  et~al.}{2016}]{2016A&A...593A.112K}
{Kraus} M.,  et~al., 2016, \mn@doi [\aap] {10.1051/0004-6361/201628493}, \href
  {http://adsabs.harvard.edu/abs/2016A%26A...593A.112K} {593, A112}

\bibitem[\protect\citeauthoryear{{Kraus}, {Arias}, {Cidale}  \&
  {Torres}}{{Kraus} et~al.}{2020}]{2020MNRAS.493.4308K}
{Kraus} M.,  {Arias} M.~L.,  {Cidale} L.~S.,   {Torres} A.~F.,  2020, \mn@doi
  [\mnras] {10.1093/mnras/staa519}, \href
  {https://ui.adsabs.harvard.edu/abs/2020MNRAS.493.4308K} {493, 4308}

\bibitem[\protect\citeauthoryear{{Kraus}, {Kourniotis}, {Arias}, {Torres}  \&
  {Nickeler}}{{Kraus} et~al.}{2023}]{2023Galax..11...76K}
{Kraus} M.,  {Kourniotis} M.,  {Arias} M.~L.,  {Torres} A.~F.,   {Nickeler}
  D.~H.,  2023, \mn@doi [Galaxies] {10.3390/galaxies11030076}, \href
  {https://ui.adsabs.harvard.edu/abs/2023Galax..11...76K} {11, 76}

\bibitem[\protect\citeauthoryear{{Kurucz}}{{Kurucz}}{1992}]{1992IAUS..149..225K}
{Kurucz} R.~L.,  1992, in {Barbuy} B.,  {Renzini} A.,  eds,  International
  Astronomical Union Vol. 149, The Stellar Populations of Galaxies. p.~225

\bibitem[\protect\citeauthoryear{{Lambert}, {Hinkle}  \& {Hall}}{{Lambert}
  et~al.}{1981}]{1981ApJ...248..638L}
{Lambert} D.~L.,  {Hinkle} K.~H.,   {Hall} D.~N.~B.,  1981, \mn@doi [\apj]
  {10.1086/159189}, \href
  {https://ui.adsabs.harvard.edu/abs/1981ApJ...248..638L} {248, 638}

\bibitem[\protect\citeauthoryear{{Lamers}, {Zickgraf}, {de Winter}, {Houziaux}
  \& {Zorec}}{{Lamers} et~al.}{1998}]{1998A&A...340..117L}
{Lamers} H. J.~G.~L.~M.,  {Zickgraf} F.-J.,  {de Winter} D.,  {Houziaux} L.,
  {Zorec} J.,  1998, \aap, \href
  {https://ui.adsabs.harvard.edu/abs/1998A&A...340..117L} {340, 117}

\bibitem[\protect\citeauthoryear{{Li}, {Gordon}, {Rothman}, {Tan}, {Hu},
  {Kassi}, {Campargue}  \& {Medvedev}}{{Li} et~al.}{2015}]{2015ApJS..216...15L}
{Li} G.,  {Gordon} I.~E.,  {Rothman} L.~S.,  {Tan} Y.,  {Hu} S.-M.,  {Kassi}
  S.,  {Campargue} A.,   {Medvedev} E.~S.,  2015, \mn@doi [\apjs]
  {10.1088/0067-0049/216/1/15}, \href
  {https://ui.adsabs.harvard.edu/abs/2015ApJS..216...15L} {216, 15}

\bibitem[\protect\citeauthoryear{{Liermann}, {Kraus}, {Schnurr}  \&
  {Fernandes}}{{Liermann} et~al.}{2010}]{2010MNRAS.408L...6L}
{Liermann} A.,  {Kraus} M.,  {Schnurr} O.,   {Fernandes} M.~B.,  2010, \mn@doi
  [\mnras] {10.1111/j.1745-3933.2010.00915.x}, \href
  {https://ui.adsabs.harvard.edu/abs/2010MNRAS.408L...6L} {408, L6}

\bibitem[\protect\citeauthoryear{{Luo}, {Zhao}, {Zhao}  \& {et al.}}{{Luo}
  et~al.}{2022}]{2022yCat.5156....0L}
{Luo} A.~L.,  {Zhao} Y.~H.,  {Zhao} G.,   {et al.} 2022, {VizieR Online Data
  Catalog: LAMOST DR7 catalogs (Luo+, 2019)}, VizieR On-line Data Catalog:
  V/156. Originally published in: 2019RAA..in.prep..L

\bibitem[\protect\citeauthoryear{{Mahy} et~al.,}{{Mahy}
  et~al.}{2022}]{2022A&A...657A...4M}
{Mahy} L.,  et~al., 2022, \mn@doi [\aap] {10.1051/0004-6361/202040062}, \href
  {https://ui.adsabs.harvard.edu/abs/2022A&A...657A...4M} {657, A4}

\bibitem[\protect\citeauthoryear{{Maravelias}, {Kraus}, {Cidale}, {Borges
  Fernandes}, {Arias}, {Cur{\'e}}  \& {Vasilopoulos}}{{Maravelias}
  et~al.}{2018}]{2018MNRAS.480..320M}
{Maravelias} G.,  {Kraus} M.,  {Cidale} L.~S.,  {Borges Fernandes} M.,  {Arias}
  M.~L.,  {Cur{\'e}} M.,   {Vasilopoulos} G.,  2018, \mn@doi [\mnras]
  {10.1093/mnras/sty1747}, \href
  {https://ui.adsabs.harvard.edu/abs/2018MNRAS.480..320M} {480, 320}

\bibitem[\protect\citeauthoryear{{Maravelias}, {Bonanos}, {Tramper}, {de Wit},
  {Yang}  \& {Bonfini}}{{Maravelias} et~al.}{2022}]{2022A&A...666A.122M}
{Maravelias} G.,  {Bonanos} A.~Z.,  {Tramper} F.,  {de Wit} S.,  {Yang} M.,
  {Bonfini} P.,  2022, \mn@doi [\aap] {10.1051/0004-6361/202141397}, \href
  {https://ui.adsabs.harvard.edu/abs/2022A&A...666A.122M} {666, A122}

\bibitem[\protect\citeauthoryear{{Maravelias}, {de Wit}, {Bonanos}, {Tramper},
  {Munoz-Sanchez}  \& {Christodoulou}}{{Maravelias}
  et~al.}{2023}]{2023Galax..11...79M}
{Maravelias} G.,  {de Wit} S.,  {Bonanos} A.~Z.,  {Tramper} F.,
  {Munoz-Sanchez} G.,   {Christodoulou} E.,  2023, \mn@doi [Galaxies]
  {10.3390/galaxies11030079}, \href
  {https://ui.adsabs.harvard.edu/abs/2023Galax..11...79M} {11, 79}

\bibitem[\protect\citeauthoryear{{Marchant} \& {Bodensteiner}}{{Marchant} \&
  {Bodensteiner}}{2024}]{2024ARA&A..62...21M}
{Marchant} P.,  {Bodensteiner} J.,  2024, \mn@doi [\araa]
  {10.1146/annurev-astro-052722-105936}, \href
  {https://ui.adsabs.harvard.edu/abs/2024ARA&A..62...21M} {62, 21}

\bibitem[\protect\citeauthoryear{{Martin} \& {Humphreys}}{{Martin} \&
  {Humphreys}}{2017}]{2017AJ....154...81M}
{Martin} J.~C.,  {Humphreys} R.~M.,  2017, \mn@doi [\aj]
  {10.3847/1538-3881/aa7e2e}, \href
  {https://ui.adsabs.harvard.edu/abs/2017AJ....154...81M} {154, 81}

\bibitem[\protect\citeauthoryear{{Martins} \& {Palacios}}{{Martins} \&
  {Palacios}}{2013}]{2013A&A...560A..16M}
{Martins} F.,  {Palacios} A.,  2013, \mn@doi [\aap]
  {10.1051/0004-6361/201322480}, \href
  {https://ui.adsabs.harvard.edu/abs/2013A&A...560A..16M} {560, A16}

\bibitem[\protect\citeauthoryear{{Masci} et~al.,}{{Masci}
  et~al.}{2019}]{2019PASP..131a8003M}
{Masci} F.~J.,  et~al., 2019, \mn@doi [\pasp] {10.1088/1538-3873/aae8ac}, \href
  {https://ui.adsabs.harvard.edu/abs/2019PASP..131a8003M} {131, 018003}

\bibitem[\protect\citeauthoryear{{Massey}, {Bianchi}, {Hutchings}  \&
  {Stecher}}{{Massey} et~al.}{1996}]{1996ApJ...469..629M}
{Massey} P.,  {Bianchi} L.,  {Hutchings} J.~B.,   {Stecher} T.~P.,  1996,
  \mn@doi [\apj] {10.1086/177811}, \href
  {https://ui.adsabs.harvard.edu/abs/1996ApJ...469..629M} {469, 629}

\bibitem[\protect\citeauthoryear{{Massey}, {Olsen}, {Hodge}, {Strong},
  {Jacoby}, {Schlingman}  \& {Smith}}{{Massey}
  et~al.}{2006}]{2006AJ....131.2478M}
{Massey} P.,  {Olsen} K.~A.~G.,  {Hodge} P.~W.,  {Strong} S.~B.,  {Jacoby}
  G.~H.,  {Schlingman} W.,   {Smith} R.~C.,  2006, \mn@doi [\aj]
  {10.1086/503256}, \href
  {https://ui.adsabs.harvard.edu/abs/2006AJ....131.2478M} {131, 2478}

\bibitem[\protect\citeauthoryear{{Massey}, {McNeill}, {Olsen}, {Hodge},
  {Blaha}, {Jacoby}, {Smith}  \& {Strong}}{{Massey}
  et~al.}{2007}]{2007AJ....134.2474M}
{Massey} P.,  {McNeill} R.~T.,  {Olsen} K.~A.~G.,  {Hodge} P.~W.,  {Blaha} C.,
  {Jacoby} G.~H.,  {Smith} R.~C.,   {Strong} S.~B.,  2007, \mn@doi [\aj]
  {10.1086/523658}, \href
  {https://ui.adsabs.harvard.edu/abs/2007AJ....134.2474M} {134, 2474}

\bibitem[\protect\citeauthoryear{{Massey}, {Neugent}, {Morrell}  \&
  {Hillier}}{{Massey} et~al.}{2014}]{2014ApJ...788...83M}
{Massey} P.,  {Neugent} K.~F.,  {Morrell} N.,   {Hillier} D.~J.,  2014, \mn@doi
  [\apj] {10.1088/0004-637X/788/1/83}, \href
  {https://ui.adsabs.harvard.edu/abs/2014ApJ...788...83M} {788, 83}

\bibitem[\protect\citeauthoryear{{Massey}, {Neugent}  \& {Smart}}{{Massey}
  et~al.}{2016}]{2016AJ....152...62M}
{Massey} P.,  {Neugent} K.~F.,   {Smart} B.~M.,  2016, \mn@doi [\aj]
  {10.3847/0004-6256/152/3/62}, \href
  {https://ui.adsabs.harvard.edu/abs/2016AJ....152...62M} {152, 62}

\bibitem[\protect\citeauthoryear{{McGregor}, {Hyland}  \& {Hillier}}{{McGregor}
  et~al.}{1988a}]{1988ApJ...324.1071M}
{McGregor} P.~J.,  {Hyland} A.~R.,   {Hillier} D.~J.,  1988a, \mn@doi [\apj]
  {10.1086/165964}, \href
  {https://ui.adsabs.harvard.edu/abs/1988ApJ...324.1071M} {324, 1071}

\bibitem[\protect\citeauthoryear{{McGregor}, {Hillier}  \& {Hyland}}{{McGregor}
  et~al.}{1988b}]{1988ApJ...334..639M}
{McGregor} P.~J.,  {Hillier} D.~J.,   {Hyland} A.~R.,  1988b, \mn@doi [\apj]
  {10.1086/166867}, \href
  {https://ui.adsabs.harvard.edu/abs/1988ApJ...334..639M} {334, 639}

\bibitem[\protect\citeauthoryear{{McGregor}, {Hyland}  \& {McGinn}}{{McGregor}
  et~al.}{1989}]{1989A&A...223..237M}
{McGregor} P.~J.,  {Hyland} A.~R.,   {McGinn} M.~T.,  1989, \aap, \href
  {https://ui.adsabs.harvard.edu/abs/1989A&A...223..237M} {223, 237}

\bibitem[\protect\citeauthoryear{{Morris}, {Eenens}, {Hanson}, {Conti}  \&
  {Blum}}{{Morris} et~al.}{1996}]{1996ApJ...470..597M}
{Morris} P.~W.,  {Eenens} P.~R.~J.,  {Hanson} M.~M.,  {Conti} P.~S.,   {Blum}
  R.~D.,  1996, \mn@doi [\apj] {10.1086/177892}, \href
  {https://ui.adsabs.harvard.edu/abs/1996ApJ...470..597M} {470, 597}

\bibitem[\protect\citeauthoryear{{Muratore}, {de Wit}, {Kraus}, {Aret},
  {Cidale}, {Borges Fernandes}, {Oudmaijer}  \& {Wheelwright}}{{Muratore}
  et~al.}{2012}]{2012ASPC..464...67M}
{Muratore} M.~F.,  {de Wit} W.~J.,  {Kraus} M.,  {Aret} A.,  {Cidale} L.~S.,
  {Borges Fernandes} M.,  {Oudmaijer} R.~D.,   {Wheelwright} H.~E.,  2012, in
  {Carciofi} A.~C.,  {Rivinius} T.,  eds,  Astronomical Society of the Pacific
  Conference Series Vol. 464, Circumstellar Dynamics at High Resolution. p.~67
  (\mn@eprint {arXiv} {1212.4798}), \mn@doi{10.48550/arXiv.1212.4798}

\bibitem[\protect\citeauthoryear{{Muratore}, {Kraus}, {Oksala}, {Arias},
  {Cidale}, {Borges Fernandes}  \& {Liermann}}{{Muratore}
  et~al.}{2015}]{2015AJ....149...13M}
{Muratore} M.~F.,  {Kraus} M.,  {Oksala} M.~E.,  {Arias} M.~L.,  {Cidale} L.,
  {Borges Fernandes} M.,   {Liermann} A.,  2015, \mn@doi [\aj]
  {10.1088/0004-6256/149/1/13}, \href
  {https://ui.adsabs.harvard.edu/abs/2015AJ....149...13M} {149, 13}

\bibitem[\protect\citeauthoryear{{Neugent}}{{Neugent}}{2021}]{2021ApJ...908...87N}
{Neugent} K.~F.,  2021, \mn@doi [\apj] {10.3847/1538-4357/abd47b}, \href
  {https://ui.adsabs.harvard.edu/abs/2021ApJ...908...87N} {908, 87}

\bibitem[\protect\citeauthoryear{{Neugent}, {Levesque}  \& {Massey}}{{Neugent}
  et~al.}{2018}]{2018AJ....156..225N}
{Neugent} K.~F.,  {Levesque} E.~M.,   {Massey} P.,  2018, \mn@doi [\aj]
  {10.3847/1538-3881/aae4e0}, \href
  {https://ui.adsabs.harvard.edu/abs/2018AJ....156..225N} {156, 225}

\bibitem[\protect\citeauthoryear{{Neugent}, {Levesque}, {Massey}  \&
  {Morrell}}{{Neugent} et~al.}{2019}]{2019ApJ...875..124N}
{Neugent} K.~F.,  {Levesque} E.~M.,  {Massey} P.,   {Morrell} N.~I.,  2019,
  \mn@doi [\apj] {10.3847/1538-4357/ab1012}, \href
  {https://ui.adsabs.harvard.edu/abs/2019ApJ...875..124N} {875, 124}

\bibitem[\protect\citeauthoryear{{Neugent}, {Levesque}, {Massey}, {Morrell}  \&
  {Drout}}{{Neugent} et~al.}{2020}]{2020ApJ...900..118N}
{Neugent} K.~F.,  {Levesque} E.~M.,  {Massey} P.,  {Morrell} N.~I.,   {Drout}
  M.~R.,  2020, \mn@doi [\apj] {10.3847/1538-4357/ababaa}, \href
  {https://ui.adsabs.harvard.edu/abs/2020ApJ...900..118N} {900, 118}

\bibitem[\protect\citeauthoryear{{Oksala}, {Kraus}, {Arias}, {Borges
  Fernandes}, {Cidale}, {Muratore}  \& {Cur{\'e}}}{{Oksala}
  et~al.}{2012}]{2012MNRAS.426L..56O}
{Oksala} M.~E.,  {Kraus} M.,  {Arias} M.~L.,  {Borges Fernandes} M.,  {Cidale}
  L.,  {Muratore} M.~F.,   {Cur{\'e}} M.,  2012, \mn@doi [\mnras]
  {10.1111/j.1745-3933.2012.01323.x}, \href
  {https://ui.adsabs.harvard.edu/abs/2012MNRAS.426L..56O} {426, L56}

\bibitem[\protect\citeauthoryear{{Oksala}, {Kraus}, {Cidale}, {Muratore}  \&
  {Borges Fernandes}}{{Oksala} et~al.}{2013}]{2013A&A...558A..17O}
{Oksala} M.~E.,  {Kraus} M.,  {Cidale} L.~S.,  {Muratore} M.~F.,   {Borges
  Fernandes} M.,  2013, \mn@doi [\aap] {10.1051/0004-6361/201321568}, \href
  {http://adsabs.harvard.edu/abs/2013A%26A...558A..17O} {558, A17}

\bibitem[\protect\citeauthoryear{{Oudmaijer} \& {de Wit}}{{Oudmaijer} \& {de
  Wit}}{2013}]{2013A&A...551A..69O}
{Oudmaijer} R.~D.,  {de Wit} W.~J.,  2013, \mn@doi [\aap]
  {10.1051/0004-6361/201220185}, \href
  {https://ui.adsabs.harvard.edu/abs/2013A&A...551A..69O} {551, A69}

\bibitem[\protect\citeauthoryear{{Oudmaijer}, {Davies}, {de Wit}  \&
  {Patel}}{{Oudmaijer} et~al.}{2009}]{2009ASPC..412...17O}
{Oudmaijer} R.~D.,  {Davies} B.,  {de Wit} W.~J.,   {Patel} M.,  2009, in
  {Luttermoser} D.~G.,  {Smith} B.~J.,   {Stencel} R.~E.,  eds,  Astronomical
  Society of the Pacific Conference Series Vol. 412, The Biggest, Baddest,
  Coolest Stars. p.~17 (\mn@eprint {arXiv} {0801.2315}),
  \mn@doi{10.48550/arXiv.0801.2315}

\bibitem[\protect\citeauthoryear{{Pantaleoni Gonz{\'a}lez}, {Ma{\'\i}z
  Apell{\'a}niz}, {Barb{\'a}}  \& {Negueruela}}{{Pantaleoni Gonz{\'a}lez}
  et~al.}{2020}]{2020RNAAS...4...12P}
{Pantaleoni Gonz{\'a}lez} M.,  {Ma{\'\i}z Apell{\'a}niz} J.,  {Barb{\'a}}
  R.~H.,   {Negueruela} I.,  2020, \mn@doi [Research Notes of the American
  Astronomical Society] {10.3847/2515-5172/ab712b}, \href
  {https://ui.adsabs.harvard.edu/abs/2020RNAAS...4...12P} {4, 12}

\bibitem[\protect\citeauthoryear{{Patrick}, {Thilker}, {Lennon}, {Bianchi},
  {Schootemeijer}, {Dorda}, {Langer}  \& {Negueruela}}{{Patrick}
  et~al.}{2022}]{2022MNRAS.513.5847P}
{Patrick} L.~R.,  {Thilker} D.,  {Lennon} D.~J.,  {Bianchi} L.,
  {Schootemeijer} A.,  {Dorda} R.,  {Langer} N.,   {Negueruela} I.,  2022,
  \mn@doi [\mnras] {10.1093/mnras/stac1139}, \href
  {https://ui.adsabs.harvard.edu/abs/2022MNRAS.513.5847P} {513, 5847}

\bibitem[\protect\citeauthoryear{{Patrick}, {Lennon}, {Schootemeijer},
  {Bianchi}, {Negueruela}, {Langer}, {Thilker}  \& {Dorda}}{{Patrick}
  et~al.}{2024}]{2024arXiv241218554P}
{Patrick} L.~R.,  {Lennon} D.~J.,  {Schootemeijer} A.,  {Bianchi} L.,
  {Negueruela} I.,  {Langer} N.,  {Thilker} D.,   {Dorda} R.,  2024, \mn@doi
  [arXiv e-prints] {10.48550/arXiv.2412.18554}, \href
  {https://ui.adsabs.harvard.edu/abs/2024arXiv241218554P} {p. arXiv:2412.18554}

\bibitem[\protect\citeauthoryear{{Peacock}, {Maccarone}, {Knigge}, {Kundu},
  {Waters}, {Zepf}  \& {Zurek}}{{Peacock} et~al.}{2010}]{2010MNRAS.402..803P}
{Peacock} M.~B.,  {Maccarone} T.~J.,  {Knigge} C.,  {Kundu} A.,  {Waters}
  C.~Z.,  {Zepf} S.~E.,   {Zurek} D.~R.,  2010, \mn@doi [\mnras]
  {10.1111/j.1365-2966.2009.15952.x}, \href
  {https://ui.adsabs.harvard.edu/abs/2010MNRAS.402..803P} {402, 803}

\bibitem[\protect\citeauthoryear{{Plez}}{{Plez}}{2012}]{2012ascl.soft05004P}
{Plez} B.,  2012, {Turbospectrum: Code for spectral synthesis}, Astrophysics
  Source Code Library, record ascl:1205.004

\bibitem[\protect\citeauthoryear{{Sana} et~al.,}{{Sana}
  et~al.}{2012}]{2012Sci...337..444S}
{Sana} H.,  et~al., 2012, \mn@doi [Science] {10.1126/science.1223344}, \href
  {https://ui.adsabs.harvard.edu/abs/2012Sci...337..444S} {337, 444}

\bibitem[\protect\citeauthoryear{{Sarkisyan} et~al.,}{{Sarkisyan}
  et~al.}{2020}]{2020MNRAS.497..687S}
{Sarkisyan} A.,  et~al., 2020, \mn@doi [\mnras] {10.1093/mnras/staa1729}, \href
  {https://ui.adsabs.harvard.edu/abs/2020MNRAS.497..687S} {497, 687}

\bibitem[\protect\citeauthoryear{{Sholukhova}, {Bizyaev}, {Fabrika},
  {Sarkisyan}, {Malanushenko}  \& {Valeev}}{{Sholukhova}
  et~al.}{2015}]{2015MNRAS.447.2459S}
{Sholukhova} O.,  {Bizyaev} D.,  {Fabrika} S.,  {Sarkisyan} A.,  {Malanushenko}
  V.,   {Valeev} A.,  2015, \mn@doi [\mnras] {10.1093/mnras/stu2597}, \href
  {https://ui.adsabs.harvard.edu/abs/2015MNRAS.447.2459S} {447, 2459}

\bibitem[\protect\citeauthoryear{{Sholukhova} et~al.,}{{Sholukhova}
  et~al.}{2024}]{2024arXiv240615270S}
{Sholukhova} O.~N.,  et~al., 2024, \mn@doi [arXiv e-prints]
  {10.48550/arXiv.2406.15270}, \href
  {https://ui.adsabs.harvard.edu/abs/2024arXiv240615270S} {p. arXiv:2406.15270}

\bibitem[\protect\citeauthoryear{{Shporer} \& {Mazeh}}{{Shporer} \&
  {Mazeh}}{2006}]{2006MNRAS.370.1429S}
{Shporer} A.,  {Mazeh} T.,  2006, \mn@doi [\mnras]
  {10.1111/j.1365-2966.2006.10554.x}, \href
  {https://ui.adsabs.harvard.edu/abs/2006MNRAS.370.1429S} {370, 1429}

\bibitem[\protect\citeauthoryear{{Skrutskie} et~al.,}{{Skrutskie}
  et~al.}{2006}]{2006AJ....131.1163S}
{Skrutskie} M.~F.,  et~al., 2006, \mn@doi [\aj] {10.1086/498708}, \href
  {https://ui.adsabs.harvard.edu/abs/2006AJ....131.1163S} {131, 1163}

\bibitem[\protect\citeauthoryear{{Solovyeva} et~al.,}{{Solovyeva}
  et~al.}{2019}]{2019MNRAS.484L..24S}
{Solovyeva} Y.,  et~al., 2019, \mn@doi [\mnras] {10.1093/mnrasl/sly241}, \href
  {https://ui.adsabs.harvard.edu/abs/2019MNRAS.484L..24S} {484, L24}

\bibitem[\protect\citeauthoryear{{Soraisam} et~al.,}{{Soraisam}
  et~al.}{2018}]{2018ApJ...859...73S}
{Soraisam} M.~D.,  et~al., 2018, \mn@doi [\apj] {10.3847/1538-4357/aabc59},
  \href {https://ui.adsabs.harvard.edu/abs/2018ApJ...859...73S} {859, 73}

\bibitem[\protect\citeauthoryear{{Sz{\'e}csi}, {Agrawal}, {W{\"u}nsch}  \&
  {Langer}}{{Sz{\'e}csi} et~al.}{2022}]{2022A&A...658A.125S}
{Sz{\'e}csi} D.,  {Agrawal} P.,  {W{\"u}nsch} R.,   {Langer} N.,  2022, \mn@doi
  [\aap] {10.1051/0004-6361/202141536}, \href
  {https://ui.adsabs.harvard.edu/abs/2022A&A...658A.125S} {658, A125}

\bibitem[\protect\citeauthoryear{{Szeifert}, {Humphreys}, {Davidson}, {Jones},
  {Stahl}, {Wolf}  \& {Zickgraf}}{{Szeifert}
  et~al.}{1996}]{1996A&A...314..131S}
{Szeifert} T.,  {Humphreys} R.~M.,  {Davidson} K.,  {Jones} T.~J.,  {Stahl} O.,
   {Wolf} B.,   {Zickgraf} F.~J.,  1996, \aap, \href
  {https://ui.adsabs.harvard.edu/abs/1996A&A...314..131S} {314, 131}

\bibitem[\protect\citeauthoryear{{Torres}, {Cidale}, {Kraus}, {Arias},
  {Barb{\'a}}, {Maravelias}  \& {Borges Fernandes}}{{Torres}
  et~al.}{2018}]{2018A&A...612A.113T}
{Torres} A.~F.,  {Cidale} L.~S.,  {Kraus} M.,  {Arias} M.~L.,  {Barb{\'a}}
  R.~H.,  {Maravelias} G.,   {Borges Fernandes} M.,  2018, \mn@doi [\aap]
  {10.1051/0004-6361/201731723}, \href
  {http://adsabs.harvard.edu/abs/2018A%26A...612A.113T} {612, A113}

\bibitem[\protect\citeauthoryear{{U}, {Urbaneja}, {Kudritzki}, {Jacobs},
  {Bresolin}  \& {Przybilla}}{{U} et~al.}{2009}]{2009ApJ...704.1120U}
{U} V.,  {Urbaneja} M.~A.,  {Kudritzki} R.-P.,  {Jacobs} B.~A.,  {Bresolin} F.,
    {Przybilla} N.,  2009, \mn@doi [\apj] {10.1088/0004-637X/704/2/1120}, \href
  {https://ui.adsabs.harvard.edu/abs/2009ApJ...704.1120U} {704, 1120}

\bibitem[\protect\citeauthoryear{{Valeev}, {Sholukhova}  \& {Fabrika}}{{Valeev}
  et~al.}{2009}]{2009MNRAS.396L..21V}
{Valeev} A.~F.,  {Sholukhova} O.,   {Fabrika} S.,  2009, \mn@doi [\mnras]
  {10.1111/j.1745-3933.2009.00654.x}, \href
  {https://ui.adsabs.harvard.edu/abs/2009MNRAS.396L..21V} {396, L21}

\bibitem[\protect\citeauthoryear{{Walborn} \& {Fitzpatrick}}{{Walborn} \&
  {Fitzpatrick}}{2000}]{2000PASP..112...50W}
{Walborn} N.~R.,  {Fitzpatrick} E.~L.,  2000, \mn@doi [\pasp] {10.1086/316490},
  \href {https://ui.adsabs.harvard.edu/abs/2000PASP..112...50W} {112, 50}

\bibitem[\protect\citeauthoryear{{Weis} \& {Bomans}}{{Weis} \&
  {Bomans}}{2020}]{2020Galax...8...20W}
{Weis} K.,  {Bomans} D.~J.,  2020, \mn@doi [Galaxies]
  {10.3390/galaxies8010020}, \href
  {https://ui.adsabs.harvard.edu/abs/2020Galax...8...20W} {8, 20}

\bibitem[\protect\citeauthoryear{{Wittkowski} et~al.,}{{Wittkowski}
  et~al.}{2017}]{2017A&A...606L...1W}
{Wittkowski} M.,  et~al., 2017, \mn@doi [\aap] {10.1051/0004-6361/201731569},
  \href {https://ui.adsabs.harvard.edu/abs/2017A&A...606L...1W} {606, L1}

\bibitem[\protect\citeauthoryear{{Yershov}}{{Yershov}}{2014}]{2014Ap&SS.354...97Y}
{Yershov} V.~N.,  2014, \mn@doi [\apss] {10.1007/s10509-014-1944-5}, \href
  {https://ui.adsabs.harvard.edu/abs/2014Ap&SS.354...97Y} {354, 97}

\bibitem[\protect\citeauthoryear{{Yusof} et~al.,}{{Yusof}
  et~al.}{2022}]{2022MNRAS.511.2814Y}
{Yusof} N.,  et~al., 2022, \mn@doi [\mnras] {10.1093/mnras/stac230}, \href
  {https://ui.adsabs.harvard.edu/abs/2022MNRAS.511.2814Y} {511, 2814}

\bibitem[\protect\citeauthoryear{{de Jager}}{{de
  Jager}}{1998}]{1998A&ARv...8..145D}
{de Jager} C.,  1998, \mn@doi [\aapr] {10.1007/s001590050009}, \href
  {https://ui.adsabs.harvard.edu/abs/1998A&ARv...8..145D} {8, 145}

\bibitem[\protect\citeauthoryear{{van den Bergh}, {Herbst}  \& {Kowal}}{{van
  den Bergh} et~al.}{1975}]{1975ApJS...29..303V}
{van den Bergh} S.,  {Herbst} E.,   {Kowal} C.~T.,  1975, \mn@doi [\apjs]
  {10.1086/190344}, \href
  {https://ui.adsabs.harvard.edu/abs/1975ApJS...29..303V} {29, 303}

\makeatother
\end{thebibliography}

%\begin{thebibliography}{99}
%\end{thebibliography}

%%%%%%%%%%%%%%%%%%%%%%%%%%%%%%%%%%%%%%%%%%%%%%%%%%

%%%%%%%%%%%%%%%%% APPENDICES %%%%%%%%%%%%%%%%%%%%%
%
%\appendix


% Don't change these lines
\bsp	% typesetting comment
\label{lastpage}
\end{document}